\documentclass[eqsecnum, nofootinbib, prd, preprintnumbers, showpacs,
superscriptaddress, twocolumn]{revtex4}

\usepackage{amsmath}
\usepackage{amssymb}
\usepackage{amsthm}
\usepackage{graphicx}
\usepackage{ifpdf}
\usepackage[latin9]{inputenc}
\usepackage{mathpazo}
\usepackage{psfrag}
\usepackage{subfigure}
\usepackage{url}

\usepackage{hyperref}

\ifpdf
\else
\usepackage{breakurl}
\fi

\def\M{\mathcal{M}}
\def\fie{\varphi}
\def\Lie{\mathcal{L}}

\def\ADM{\textrm{\mbox{\tiny{ADM}}}}
\def\NS{\textrm{\mbox{\tiny{NS}}}}

\def\Kphy{\bar{K}}
\def\hphy{\bar{q}}

\def\Fm{\mathcal{F}^{(m)}}
\def\Fg{\mathcal{F}^{(g)}}
\def\Jm{\mathcal{J}^{(m)}}
\def\Jg{\mathcal{J}^{(g)}}
\def\R{\mathcal{R}}
\def\ls{{(\ell)}}
\def\ns{{(n)}}
\def\grad{\nabla}

\theoremstyle{definition}

\begin{document}

\title{Introduction to dynamical horizons in numerical relativity}

\author{Erik Schnetter}
\email{schnetter@cct.lsu.edu}

\affiliation{Center for Computation and Technology,
302 Johnston Hall, Louisiana State University, Baton Rouge, LA 70803, USA}
\homepage{http://www.cct.lsu.edu/about/focus/numerical/}

\affiliation{Max-Planck-Institut für Gravitationsphysik,
Albert-Einstein-Institut, Am Mühlenberg 1, D-14476 Golm, Germany}
\homepage{http://numrel.aei.mpg.de/}
\homepage{http://www.aei.mpg.de/}

\author{Badri Krishnan}
\email{badri.krishnan@aei.mpg.de}

\affiliation{Max-Planck-Institut für Gravitationsphysik,
Albert-Einstein-Institut, Am Mühlenberg 1, D-14476 Golm, Germany}

\author{Florian Beyer}
\email{florian.beyer@aei.mpg.de}

\affiliation{Max-Planck-Institut für Gravitationsphysik,
Albert-Einstein-Institut, Am Mühlenberg 1, D-14476 Golm, Germany}

\date{May 31, 2006}

    
\begin{abstract}

  This paper presents a quasi-local method of studying the physics of
  dynamical black holes in numerical simulations.  This is done within
  the dynamical horizon framework, which extends the earlier work on
  isolated horizons to time-dependent situations.  In particular: (i)
  We locate various kinds of marginal surfaces and study their time
  evolution. An important ingredient is the calculation of the
  signature of the horizon, which can be either spacelike, timelike,
  or null. (ii) We generalize the calculation of the black hole mass
  and angular momentum, which were previously defined for axisymmetric
  isolated horizons to dynamical situations. (iii) We calculate the
  source multipole moments of the black hole which can be used to
  verify that the black hole settles down to a Kerr solution. (iv) We
  also study the fluxes of energy crossing the horizon, which
  describes how a black hole grows as it accretes matter and/or
  radiation.

  We describe our numerical implementation of these concepts and apply
  them to three specific test cases, namely, the axisymmetric head-on
  collision of two black holes, the axisymmetric collapse of a neutron
  star, and a non-axisymmetric black hole collision with non-zero
  initial orbital angular momentum.

\end{abstract}


\preprint{AEI-2006-018, LSU-REL-033006}

\maketitle


\section{Introduction}
\label{sec:intro}

In spite of fundamental advances in our understanding of black holes,
relatively little is known about them in the fully non-perturbative,
dynamical regime of general relativity.  Most of our intuition
regarding black holes comes from studying the stationary, axisymmetric
Kerr-Newman solutions, and perturbations thereof. This, along with
post-Newtonian calculations which treat the black hole as a point
particle, are usually adequate for understanding many astrophysical
processes involving black holes. However, understanding the
gravitational waveforms arising due to, say, the merger phase of the
coalescence of two black holes or the gravitational collapse of a
star, will require us to go beyond perturbation theory and to confront
the non-linearities and dynamics of the full Einstein equations. This
regime may contain qualitatively new, non-perturbative features. In
this paper, we discuss an important ingredient for understanding this
regime, namely, the dynamics of the black hole horizon.  Numerical
simulations of black holes have greatly improved in the last few
years.  Simulations of the entire merger process, starting from the
last few orbits of the inspiral right up to the ringdown have become
possible in the past year \cite{Pretorius:2005gq, Campanelli:2005dd,
  Baker05a, Diener-etal-2006a, Herrmann:2006ks, Campanelli:2006gf,
  Baker:2006yw, Campanelli:2006uy}.
It is then important to look for better ways to
extract more physical information from simulations and to compare
results from two different simulations performed using different
coordinate systems, gauge conditions etc. This can be a non-trivial
task in itself, and understanding black hole horizons is a necessary
ingredient.

Due to their global nature, black hole event horizons can only be
located once a simulation is complete and we have obtained the full
spacetime.  In numerical simulations, it is instead common to use
marginally trapped surfaces to locate black holes on a Cauchy surface
in real time.  We use the formalism of dynamical horizons
\cite{Ashtekar02a,Ashtekar03a} to study black holes.  Using
isolated/dynamical horizons, it is shown that marginally trapped
surfaces, while not a substitute for event horizons, do have many
useful properties and can be used fruitfully to study black hole
physics.  Dynamical horizons are a significant extension of the
isolated horizon framework
\cite{Ashtekar98a,Ashtekar99a,Ashtekar01a,Ashtekar00b,Ashtekar00a},
which models isolated stationary black holes in an otherwise dynamical
spacetime.  Both these frameworks are, in turn, very closely related
to and motivated by the earlier work on trapping horizons by Hayward
\cite{Hayward94a,Hayward94b,Hayward04}. See \cite{Ashtekar:2004cn,
  Booth-review, Gourgoulhon-Jaramillo-Review} for reviews.
Information obtained from these quasi-local horizons complements the
information obtained from the event horizon.  Once a simulation is
complete and ready for post-processing, event horizons are useful for
studying global properties and the causal structure of the spacetime,
and also phenomena such as the topology change of the horizon during a
black hole coalescence. Reliable and computationally efficient codes
are now available for locating event horizons (see e.g.\
\cite{Diener03a}).  Such information cannot be obtained at the
quasi-local level, which is instead better for tracking the physical
parameters and geometry of a black hole in real time.

The dynamics of apparent and event horizons have been numerically
studied in the past in detail in axisymmetry (see e.g.\
\cite{Hughes94a, Anninos93a, Anninos94f, Brandt94c, Libson94a,
  Masso95a}).  We want to extend this work to non-axisymmetric and
non-vacuum spacetimes, and we want to emphasise non-gauge-dependent
analysis methods.  In particular, we consider the following
applications: (i) We study the behavior of various marginally trapped
surfaces under time evolution.  This leads to greater insights about
the trapped region of a spacetime.  An important ingredient here is
the signature of the world tube of marginally trapped surfaces.  This
world tube is known to be null for isolated horizons, and more
generally, it can be either spacelike or timelike; we show that both
types occur frequently in numerical simulations.  (ii) We give
meaningful definitions for the angular momentum, mass, and higher
multipole moments for the dynamical black hole.  The multipole moments
capture gauge invariant geometrical information regarding the horizon
geometry, and should be useful for understanding fundamental issues
such as the final state of black hole collapse. For example, we would
expect that after a black hole has formed and settled down, its
multipole moments should be identical to the source multipoles of a
Kerr black hole. We show that it is, in principle, possible to verify
this conjecture and to calculate the rate at which a black hole
approaches equilibrium.  (iii) We also describe and implement methods
for calculating the energy flux falling into the horizon. This gives
us detailed information on how black holes grow as they swallow matter
and radiation.

This paper is organized as follows.  Section \ref{sec:basics} sets up
notation, and summarizes the basic definitions and properties of
trapped surfaces and dynamical horizons.  Section \ref{sec:numerics}
describes the various physical quantities that we calculate using
dynamical horizons, and also their numerical implementation.  Section
\ref{sec:example} presents three concrete, well known numerical
examples where these concepts are applied and finally, section
\ref{sec:conclusion} discusses some open issues and directions for
further work.  Unless mentioned otherwise, we use geometrical units
with $G=c=1$, the spacetime signature is $(-,+,+,+)$, all manifolds
and fields are assumed to be smooth, and the Penrose abstract index
notation is used throughout.  The derivative operator compatible with
the spacetime metric $g_{ab}$ is $\grad_a$ and, following Wald
\cite{Wald84}, the Riemann tensor is defined via $(\grad_a\grad_b -
\grad_b\grad_a) \omega_c = {R_{abc}}^d\omega_d$.

\section{Basic notions and definitions}
\label{sec:basics}

\subsection{Trapped surfaces and apparent horizons}
\label{subsec:ah}

Let $S$ be a closed, orientable spacelike 2-surface in a 4-dimensional
spacetime $(\M,g_{ab})$.  The expansion of any such surface can be
defined invariantly without any reference to a time slicing of the
spacetime.  Since $S$ is smooth, spacelike, and 2-dimensional, the set
of vectors orthogonal to it at any point form a 2-dimensional
Minkowskian vector space.  Thus, we can define two linearly
independent, future-directed, null vectors $\ell^a$ and $n^a$
orthogonal to $S$ such that
\begin{eqnarray} \label{eq:normalization} g_{ab} \ell^an^b = -1 \,.\end{eqnarray}
Note that this convention is different from that used in
\cite{Ashtekar03a}. We shall assume that we know \textit{a priori} what
the outgoing and ingoing directions on $\M$ are. By convention,
$\ell^a$ will denote an outgoing null normal and $n^a$ an ingoing one.
The null normals are specified only up to a boost transformation
\begin{equation}\label{eq:boost}
\ell^a \rightarrow f\ell^a \,, \qquad n^a \rightarrow f^{-1}n^a
\end{equation}
where $f$ is a, positive definite, smooth function on $S$.  All
physical quantities must be invariant under this gauge transformation.

The Riemannian 2-metric $\tilde{q}_{ab}$ on $S$ induced by the
spacetime metric $g_{ab}$ is
\begin{equation}
\tilde{q}_{ab} = g_{ab} + \ell_an_b + n_a \ell_b\,.
\end{equation}
The tensor ${\tilde{q}_a}^b$ can be viewed as a projection operator on
to $S$. The null expansions are
\begin{equation}
\Theta_\ls = \tilde{q}^{ab}\grad_a\ell_b \,, \qquad \Theta_\ns =
\tilde{q}^{ab}\grad_an_b\,. 
\end{equation}
These expansions tell us how the area element of $S$ changes as it is
deformed along $\ell^a$ and $n^a$ respectively.

The shear of $\ell^a$, $\sigma_{\ls ab}$, is the symmetric trace-free part
of the projection of $\grad_a\ell_b$:
\begin{equation}
\sigma_{\ls ab} = \tilde{q}_a^c\tilde{q}_b^d\nabla_{(c}\ell_{d)} -
\frac{1}{2}\Theta_\ls\tilde{q}_{ab} \,.
\end{equation}
Similarly, the shear of $n^a$ is
\begin{equation}
\sigma_{\ns {ab}} = \tilde{q}_a^c\tilde{q}_b^d\nabla_{(c}n_{d)} -
\frac{1}{2}\Theta_\ns\tilde{q}_{ab} \,.
\end{equation}
Note that these definitions only involve derivatives tangential to
$S$.  Thus $\ell^a$ and $n^a$ can, if necessary, be extended
arbitrarily away from $S$ while computing these quantities.  

The closed 2-surface $S$ is said to be a \textit{trapped surface} if
both expansions $\Theta_\ls$ and $\Theta_\ns$ are strictly negative.
This is very different from a sphere in normal flat space which has
positive outgoing expansion and negative ingoing expansion.  This
definition was first introduced by Penrose \cite{Penrose65}, who
recognized its importance in the formation of singularities. On a
\emph{marginal surface}, one of the two null expansions vanish.  Of
particular interest are the \emph{marginally outer trapped surfaces}
(MOTSs), for which the outgoing null rays along $\ell^a$ have zero
expansion.  In addition, we shall mostly deal with \emph{future
  marginally outer trapped surfaces} (FMOTSs), i.e., MOTSs with
$\Theta_\ns < 0$.

There are three main reasons why closed trapped surfaces are important
for studying black holes. First, the existence of a trapped surface
implies the existence of a singularity in the future
\cite{Penrose65,Penrose70a}.  Secondly, they are guaranteed to always
lie within the event horizon.  Finally, in stationary spacetimes, the
null generators of the event horizon have zero expansion.  Thus for
stationary spacetimes, the cross-section of the event horizon is a
MOTS.

While trapped and marginally outer trapped surfaces are defined in the
full four dimensional spacetime, in numerical relativity, one usually
considers trapped surfaces in conjunction with a foliation of
(partial) Cauchy 
surfaces containing $S$; it is numerically much easier to look for
closed surfaces on the Cauchy surface rather than in the full
spacetime manifold.  For concreteness, we shall work in the ADM
formalism where the relevant portion of spacetime is foliated by
spacelike surfaces, and $\Sigma$ shall denote one of the leaves of
this foliation.  However, it will be obvious that the formalism is
applicable no matter how Einstein's equations are implemented.

The trapped region $\mathcal{T}_\Sigma$ on $\Sigma$ is defined to be
the set of points in $\Sigma$ through which there passes a trapped
surface contained entirely in $\Sigma$.  Note that there could be
points in $\Sigma$ not contained in $\mathcal{T}_\Sigma$, but through
which there passes a trapped surface not contained in $\Sigma$.  Thus,
$\mathcal{T}_\Sigma$ is a subset of the intersection of $\Sigma$ with
the 4-dimensional trapped region in the full spacetime.  A connected
component of the boundary of $\mathcal{T}_\Sigma$ is called an
\emph{apparent horizon} (AH). Under suitable regularity conditions,
the AH can be shown to be a MOTS \cite{Hawking73a, Kriele97}.  Thus,
an apparent horizon is the outermost MOTS on $\Sigma$.  Due to this
``outermost'' property, an AH is not a quasi-local object on $\Sigma$.
The behavior of AHs under time evolution can be quite irregular.  For
example, they can ``jump'' discontinuously. On the other hand, as we
shall soon see, MOTSs are more regular.

\subsection{Dynamical horizons}
\label{sec:dh}

\subsubsection{Definition and examples}
\label{subsubsec:dhdefn}

We can use marginal surfaces to extract physically interesting
information about the black hole.  The key idea is to look not at a
single MOTS by itself, but rather a \emph{world tube} $H$ of MOTSs
constructed by stacking up the MOTSs obtained by time evolution.  Such
a world tube is called a \emph{Marginally Trapped Tube} (MTT).  An MTT
is thus a smooth 3-surface foliated by MOTSs.

\emph{The existence of MTTs:} Numerically, it has been observed that
marginal surfaces (though not apparent horizons --- see below) usually
behave smoothly under time evolution and produce a smooth MTT.  This
observation is placed on a more rigorous footing by the recent result
of Andersson et al.\ \cite{ams05}, which proves the local existence of
MTTs for a large class of MOTSs.  Their results require the MOTS to be
\emph{strictly-stably-outermost}.  An MOTS $S$ on $\Sigma$ is said to
be strictly-stably-outermost if there exists an infinitesimal first
order outward deformation which makes $S$ strictly untrapped. Working
with a radial coordinate $r$ on $\Sigma$ such that $S$ is a level set
of $r$, and $r$ increases in the outward direction, a sufficient (but
not necessary) condition for $S$ to be strictly-stably-outermost is
$\partial_r \Theta_\ls(r) > 0$ everywhere\footnote{More precisely,
  $\partial_r \Theta_\ls(r) \geq 0$ with $\partial_r \Theta_\ls(r) >
  0$ somewhere on $S$.} on $S$.  Here it is understood that we obtain
$\Theta_\ls$ as a function of $r$ by calculating $\Theta_\ls$ for the
constant-$r$ surfaces in the vicinity of $S$.  In principle, for an
unfortunate choice of $r$, it might happen that $\partial_r \Theta_\ls
< 0$ even though there is a different choice for which this condition
is satisfied.  In any case, this is sufficient for verifying that $S$
is strictly-stably-outermost.\footnote{It is harder to show that a
  MOTS is \emph{not} strictly-stably-outermost. This can be done by
  calculating the signature of the horizon (see below) or by
  calculating the principle eigenvalue of the stability operator
  defined in \cite{ams05}.}  This condition, unlike the outermost
condition for an AH, is a quasi-local condition. We have found in our
simulations that most physically interesting MOTSs, such as ones which
asymptote to the event horizon, and also AHs, satisfy this condition
quite generally.  However, as we shall see, there exist also MOTSs
which are not strictly-stably-outermost. In practice, instead of
checking $\partial_r \Theta_\ls > 0$ directly, we look for a surface
with a small positive (or negative) non-vanishing expansion, and check
that it lies completely outside (or inside) the MOTS.

It is shown in \cite{ams05} that if a MOTS $S$ is
strictly-stably-outermost, then at least locally in time, $S$ is a
cross-section of a smooth MTT.  More explicitly, this result shows
that given a foliation of the spacetime by Cauchy surfaces $\Sigma_t$,
if there is a MOTS $S_0$ on $\Sigma_0$ which is
strictly-stably-outermost, then MOTSs $S_t$ exist on $\Sigma_t$ for
$-\epsilon < t < \epsilon$ (for sufficiently small $\epsilon$) such
that the union $\bigcup S_t$ is a smooth MTT. The MTT will exist for
at least as long as the MOTS remains strictly-stably-outermost. This
is a conceptually important result for numerical relativity because it
shows that a large class of MOTSs behave regularly under time
evolution.  How is this to be reconciled with the known fact that AHs
can ``jump'' during a time evolution?  The reason is simply because of
the outermost property.  It is possible that a new MOTS can appear on
the outside of a given MOTS.  The ``old'' MOTS is then no longer the
globally outermost one even though it is locally outermost, and it
continues to evolve in a perfectly regular manner, but it is no longer
an AH.

There are, as yet, no similar existence proofs for MOTSs which are not
strictly-stably-outermost.  However, as we shall see later, we find in
all the examples we have looked at, that MOTSs evolve smoothly even in
this case, forming a regular world tube.

\emph{Isolated and dynamical horizons:} An MTT is null in equilibrium
situations when no matter or radiation is falling into it; the rest of
the spacetime is still allowed to be highly dynamical.  This situation
is formalized by the notion of an isolated horizon~\cite{Ashtekar98a,
  Ashtekar99a, Ashtekar00b, Ashtekar01a, Ashtekar00a}.  Using isolated
horizons, it has been possible to derive the laws of black hole
mechanics, use it as a basis for the quantum black hole entropy
calculations and find unexpected properties of hairy black holes in
Einstein-Yang-Mills theory; see \cite{Ashtekar:2004cn} and references
therein.  Most importantly for our purposes, isolated horizons have
also proved to be useful in numerical relativity.  For example,
isolated horizons provide a coordinate invariant method of calculating
the angular momentum and mass of a black hole \cite{Dreyer02a}.  They
can be used to obtain boundary conditions for constructing
quasi-equilibrium initial data sets \cite{Dain-2005, Jaramillo:2004uc,
  Cook:2004kt, Cook-2006}.  They might have a role in waveform
extraction \cite{Ashtekar00a}. A pedagogical review of isolated
horizons from the numerical relativity perspective can be found in
\cite{Gourgoulhon-Jaramillo-Review}.

In this paper, we are more interested in the dynamical regime when the
MTT is not null.  A spacelike MTT consisting of future-marginally
trapped surfaces is called a \emph{Dynamical Horizon} (DH).  Thus, a
dynamical horizon is a spacelike 3-surface equipped with a given
foliation by FMOTSs.  The properties of a
dynamical horizon are studied in detail in \cite{Ashtekar02a,
  Ashtekar03a, Ashtekar05}. The case when the horizon is very close to
being isolated but still evolving dynamically has been studied in
\cite{Booth04a, Kavanagh06} and its Hamiltonian treatment is
considered in \cite{Booth05}.  Note that the local existence of DHs
follows from the local existence of MTTs because if $\Theta_\ns < 0$
at any given time, it will continue to be strictly negative for at
least a short duration. We elaborate on the spacelike property below.

A timelike MTT will be called a \emph{timelike membrane} (TLM).  A TLM
cannot be considered to represent the surface of a black hole since a
time-like surface is not a one-way membrane, and both ingoing and
outgoing causal curves can pass through it.  In some instances, we
shall use the term ``horizon'' loosely to refer to a generic marginal
surface or a MTT without any further qualifiers.  The exact meaning
should hopefully be clear from the context.  

\begin{figure} 
  \begin{center}
  \psfrag{taua}{$\hat{\tau}^a$}
  \psfrag{S1}{$S_1$}
  \psfrag{S2}{$S_2$}
  \psfrag{na}{$n^a$}
  \psfrag{rha}{$\hat{r}^a$}
  \psfrag{la}{$\ell^a$}
  \psfrag{DH}{$H$}
  \psfrag{S0}{$S$}
  \psfrag{Ta}{$T_a$}
  \psfrag{Ra}{$R_a$}
  \psfrag{M}{$\Sigma$}
  \includegraphics[height=3cm]{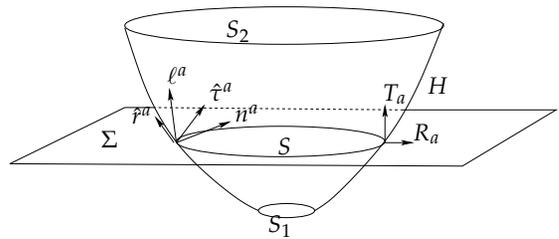}
  \caption{A dynamical horizon $H$ bounded by MOTSs $S_1$ and $S_2$.
    $\ell^a$ is the outgoing null normal, $n^a$ is the ingoing null
    normal, $\hat{r}^a$ is the unit spacelike normal to the
    cross-sections, and $\hat{\tau}^a$ is the unit timelike normal to
    $H$.  $\Sigma$ is a Cauchy surface intersecting $H$ in a 2-sphere
    $S$. $T^a$ is the unit timelike normal to $\Sigma$ and $R^a$ is the
    unit space-like outward pointing vector normal to $S$ and tangent to
    $\Sigma$. } \label{fig:cauchy}
\end{center}
\end{figure}

An explicit example of a dynamical horizon is provided by the Vaidya
spacetime which describes the gravitational collapse of null dust
\cite{Vaidya51a,Kuroda84a,Schnetter-Krishnan-2005}.  (See also
\cite{Booth05a} for further examples in spherically symmetry).  More
generally, figure \ref{fig:cauchy} depicts a dynamical horizon $H$
bounded by two MOTSs $S_1$ and $S_2$.  $S$ is a typical member of the
foliation.  The vector $\hat{\tau}^a$ is the future directed unit
timelike normal to $H$, $\hat{r}^a$ is tangent to $H$ and is the unit
outward pointing spacelike normal to the cross-sections.  A fiducial
set of null normals is
\begin{equation}\label{eq:fiducialnormals}
\ell^a = \frac{1}{\sqrt{2}}(\hat{\tau}^a + \hat{r}^a)\,, \qquad
n^a = \frac{1}{\sqrt{2}}(\hat{\tau}^a-\hat{r}^a) \,.
\end{equation}
As before, $\Theta_\ls = 0$ and $\Theta_\ns < 0$.  The area of a
cross-section $S$ will be denoted by $A_S$ and its radius by $R_S
:=\sqrt{A_S/4\pi}$.  A radial coordinate on $H$ will be denoted by
$r$; the cross sections of $H$ are the constant $r$ surfaces.  The
3-metric and extrinsic curvature of $H$ will be denoted respectively
by $q_{ab}$ and $K_{ab}$, and $\tilde{q}_{ab}$ is the 2-metric on $S$.

Figure \ref{fig:cauchy} shows also a Cauchy surface $\Sigma$
intersecting a
dynamical horizon $H$.  This intersection $S$ will always be assumed
to be one of the given cross-sections of $H$.  The unit timelike
normal to the horizon is $T^a$ and the unit outward pointing spacelike
normal to $S$ within $\Sigma$ is $R^a$.  The three metric and
extrinsic curvature of $\Sigma$ are denoted by $\bar{q}_{ab}$ and
$\bar{K}_{ab}$ respectively.  The fiducial set of null normals to $S$
arising naturally from $\Sigma$ are
\begin{equation}\label{eq:nullnormals}
\bar{\ell}^a = \frac{1}{\sqrt{2}}(T^a + R^a)\,, \qquad
\bar{n}^a = \frac{1}{\sqrt{2}}(T^a-R^a) \,.
\end{equation}
A boost transformation of the form of equation (\ref{eq:boost})
connects $(\ell^a,n^a)$ and $(\bar{\ell}^a,\bar{n}^a)$:
\begin{equation}
\ell^a = f\bar{\ell}^a\,, \qquad n^a = f^{-1}\bar{n}^a\,.
\end{equation}
When the horizon settles down and becomes null, an infinite boost
($f\rightarrow \infty$) is required to go from
$(\bar{\ell}^a,\bar{n}^a)$ to $({\ell}^a,{n}^a)$.

We conclude this sub-section with a short summary of some basic
properties of a dynamical horizon:
\begin{description}
\item[Topology:] The cross-sections of a DH can be either spherical or
  toroidal \cite{Hayward94a,Ashtekar02a,Ashtekar03a,ams05}.  Toroidal
  topology is possible only in exceptional cases when $\sigma_{\ls
    ab}$, the scalar curvature $\tilde{R}$ of $S$,
  $\Lie_\ell\Theta_\ls$, $R_{ab}\ell^b$, and $\zeta^a$ (defined in
  section \ref{sec:numerics}) all vanish on $S$ \cite{Ashtekar03a}.
  We shall therefore always take the cross-sections to be spherical.
  There are no similar results for cross-sections of TLMs. However, we
  use an apparent horizon tracker which can only locate spherical AHs
  \cite{Thornburg2003:AH-finding} and therefore all observed MOTSs
  have spherical topology.

\item[Second Law:] The area of the cross-sections of a DH increases
  along $\hat{r}^a$ \cite{Ashtekar02a,Ashtekar03a}.  Thus, if we
  choose a time evolution vector field $t^a$ for which $t\cdot\hat{r}
  > 0$, then the area of the dynamical horizon will increase in time,
  and this result can be called the second law for dynamical horizons.
  Similarly, the area of a TLM \emph{decreases} if $\Theta_\ns < 0$,
  and increases if $\Theta_\ns > 0$.

\item[Foliation and Uniqueness:] Any given spacelike MTT cannot have
  more than one distinct dynamical horizon structure on it
  \cite{Ashtekar05}.  This means that a DH can have one, and only one
  foliation by FMOTSs.  This further implies that if a Cauchy surface
  $\Sigma$ does not intersect a given DH in one of the preferred
  cross-sections, then the intersection cannot be a MOTS at all.
  Thus, different choices of Cauchy surfaces in general lead to
  different dynamical horizons.  There are however some constraints on
  the location of dynamical horizons and trapped surfaces as proved by
  Ashtekar and Galloway \cite{Ashtekar05}.  For example, they show
  that given a dynamical horizon $H$ (along with a mild genericity
  assumption), there cannot be any trapped surfaces (and therefore no
  DHs) contained entirely in the past domain of dependence of $H$.
  See also \cite{Schnetter-Krishnan-2005,Eardley98} for further
  discussion.
\end{description}

\subsubsection{The signature of a MTT}
\label{subsubsec:signature1}

As discussed above, MTTs have been shown to exist for a large and
physically interesting class of MOTSs, and this is borne out in a
large number of numerical simulations where MOTSs are located and
evolved smoothly.  How many of these MTTs are actually dynamical
horizons? In other words, when is a MTT spacelike?  The first
result in this direction was obtained by Hayward \cite{Hayward94a}
(see also \cite{Dreyer02a}).  Using the Raychaudhuri equation for
$\ell^a$, it can be shown that an MTT is spacelike if $\alpha < 0$,
null if $\alpha=0$ and timelike if $\alpha>0$, where  
\begin{equation}\label{eq:signature}
\alpha \equiv \frac{ \sigma_{\ls ab}\sigma_\ls^{ab} +
R_{ab}\ell^a\ell^b}{\Lie_n\Theta_\ls}\,.
\end{equation}
In writing this expression, it is assumed that $\ell^a$ and $n^a$ are
extended off $H$ geodetically, so that $\Lie_n\Theta_\ls$ is
meaningful.  The term in the numerator is strictly positive in the
case of dynamical horizons if the matter fields satisfy, say, the null
energy condition. It vanishes for isolated horizons.  The denominator
is negative for the Vaidya spacetime and also for the stationary
Kerr-Newman family.  This captures the notion that as we go inside the
black hole, the outgoing null rays become more and more converging.
Assuming that the numerator of Eq.~(\ref{eq:signature}) is nowhere
vanishing on $H$, the hypothesis that $H$ is spacelike is equivalent
to $\Lie_n\Theta_\ls < 0$.  As shown by Ben-Dov \cite{BenDov04a}, this
last condition is not satisfied for all MTTs; in Oppenheimer-Snyder
collapse \cite{Oppenheimer39a}, there exists a timelike world tube of
FMOTSs with $\Lie_n\Theta_\ls > 0$.

The issue of the signature has been considered in \cite{ams05}.  There
it is shown that if a MOTS $S$ is strictly stably outermost, and if
the quantity $\sigma_{\ls ab}\sigma_\ls^{ab} + R_{ab}\ell^a\ell^b$ is
non-zero \emph{somewhere} on $S$ (and assuming the null energy
condition), then the MTT containing $S$ is spacelike in a neighborhood
of $S$.  This result is stronger than Hayward's result
(Eq.~(\ref{eq:signature})) and it shows clearly that the spacelike
case is physically the most interesting because $\sigma_{\ls
  ab}\sigma_\ls^{ab} + R_{ab}\ell^a\ell^b$ will not vanish in a
non-stationary situation.  It also shows, somewhat surprisingly, that
even if matter or radiation is falling into a black hole only in the
form of say, a single narrow beam from a particular direction, the
\emph{entire} MTT is spacelike. One might naively have thought that
the MTT would be spacelike only on portions where the energy flux is
non-zero, and null otherwise.  This is not the case because of the
elliptic nature of the equations governing the deformations of a MOTS.
\footnote{The Oppenheimer-Snyder case studied in \cite{BenDov04a} does not
satisfy the hypotheses of these theorems because for this case,
the matter fields have a discontinuity at the surface of the star.
Further examples in spherical symmetry are studied in \cite{Booth05a}
where the matter fields are smooth.}

In all the examples we present later, it turns out that MOTSs form in
pairs, i.e., just after a MOTS $S_0$ appears initially, it bifurcates
into ``outer'' and ``inner'' MTTs, $H_{out}$ and $H_{in}$
respectively.  The initial MOTS $S_0$ is the common cross section of
$H_{out}$ and $H_{in}$, and the union $H_{tot} = H_{out}\bigcup
H_{in}$ forms a single smooth manifold, as far as we can tell
numerically (though a more detailed analysis of the differentiability
of $H_{tot}$ is required).  In particular, the area of the
cross-sections is a differentiable and monotonic function on this
manifold.  Furthermore, $H_{out}$ is spacelike, even on the initial
cross-section $S_0$.  This means that the inner MTT $H_{in}$ is, by
continuity, initially spacelike in an open neighborhood of $S_0$.
However, in some cases $H_{in}$ soon acquires a mixed signature and
becomes more and more timelike, and ends up as a TLM.  We strongly
suspect that such a bifurcation is a general phenomenon whenever a new
MOTS is formed.  The MOTSs on the inner MTT are not
strictly-stably-outermost and thus $H_{in}$ is not required to be
spacelike according to the results of \cite{ams05}.

\begin{figure} 
  \unitlength=1in
  \begin{center}
    \psfrag{S1}{$S_{out}$}
    \psfrag{S2}{$S_{in}$}
    \psfrag{S3}{$S_{(1)}$}
    \psfrag{S4}{$S_{(2)}$}
    \psfrag{R}{$r^a$} 
    \includegraphics[height=2.0in]{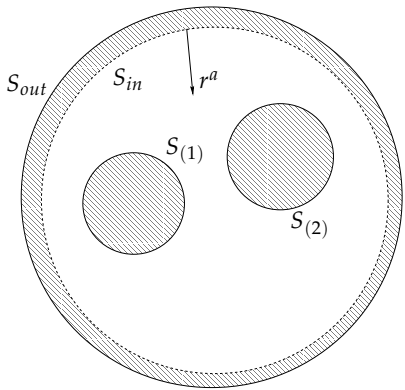}
  \end{center}
  \caption{Two MOTSs $S_{(1)}$ and $S_{(2)}$ surrounded by a common
  MOTS $S_{out}$.  Spheres lying just inside these FMOTSs must have
  negative outgoing expansion.  Thus, there must be a inner trapped
  horizon $S_{in}$ inside $S_{out}$ which encloses $S_{(1)}$ and
  $S_{(2)}$.}\label{fig:multibh} 
\end{figure}

There is one configuration where the existence of an inner MTT is plausible.
Figure \ref{fig:multibh} shows two MOTSs $S_{(1),(2)}$ surrounded by a
common MOTS $S_{out}$; $\Theta_\ls$ vanishes on all these surfaces.
Let us assume that $S_{(1)}$, $S_{(2)}$, and $S_{out}$ are all
strictly-stably-outermost and that deforming $S_{(1)}$ and $S_{(2)}$
outward yields strictly untrapped surfaces $S_{(1)}^\prime$ and
$S_{(2)}^\prime$.  Similarly, suppose that deforming $S_{out}$
\emph{inwards} gives a strictly \emph{trapped} surface
$S_{out}^\prime$.  Then, since $\Theta_\ls$ must change sign somewhere
between $S_{out}^\prime$ and $S_{(1)}^\prime$ or $S_{(2)}^\prime$, it
is plausible that there is a MOTS $S_{in}$ in the intermediate region
inside $S_{out}$ and outside $S_1$ and $S_2$.  This argument is
supported by a recent result by Schoen \cite{Schoen04} which shows the
existence of a MOTS between a trapped (in our case $S_{out}^\prime$)
and an untrapped surface (in our case $S_{(1)}^\prime \bigcup
S_{(2)}^\prime$).  It might be possible to extend this proof to
rigorously prove the existence of $S_{in}$ in our case, and to check
whether it is topologically a sphere.  $S_{(1)}$, $S_{(2)}$, and
$S_{out}$ are cross sections of a dynamical horizon while $S_{in}$ is
a cross-section of an MTT, not necessarily a dynamical horizon.

\section{Applications}
\label{sec:numerics}

This section discusses some possible applications of dynamical
horizons.  These ideas are illustrated using concrete numerical
examples later in Section \ref{sec:example}.

\subsection{Computing the signature of a MTT}
\label{subsec:signature}

From a numerical standpoint, it is more convenient to deduce the
signature of $H$ by directly calculating the induced metric $q_{ab}$,
rather than from Eq.~(\ref{eq:signature}) by calculating
$\Lie_n\Theta_\ls$ which requires extensions of $\ell^a$ and $n^a$
away from the horizon. The signature of $H$ is then determined by the
sign of the determinant of $q_{ab}$ which is gauge independent; note
that the determinant is itself gauge dependent.  To calculate $q_{ab}$
we find a frame $\mathbf{e}_{(i)}^{a}$ ($i=1,2,3$) on $H$, i.e., three
smooth vector fields on $H$ which are pointwise linearly independent.
We then simply need to compute the determinant of the matrix 
\begin{equation}
\label{eq:compute_qij}
\mathbf{q}_{(i)(j)} := g_{ab}\mathbf{e}_{(i)}^a\mathbf{e}_{(j)}^a \,.
\end{equation}

We construct a frame on $H$ as follows. Let $(t,x^i)$ ($i=1,2,3$) be
the spacetime coordinates on $\M$ used in the numerical simulation.
The MTT $H$ is topologically $I\times S^2$ ($I$ some interval in
$\mathbb R$) so that we can assume coordinates $(r,\theta,\phi)$ on
it. Here $(\theta,\phi)$ are standard coordinates on $S^2$ and $r$ is
a radial coordinate.  We can use the time coordinate $t$ as the radial
coordinate $r$ on $H$ by considering $H$ to be embedded into the
spacetime $\M$ by means of the map
\begin{equation} 
F(r,\theta,\phi)=(t=r,x^i=F^i(r,\theta,\phi)) \,.
\end{equation} 
The maps $F^i$ are known as soon as the MOTSs are found by the AH
tracker.  As a frame on $H$ we choose
\begin{equation}
\label{eq:frame_H}
\mathbf{e}_{(1)}=\partial_\theta,\quad 
\mathbf{e}_{(2)}=\frac 1{\sin\theta}\partial_\phi,\quad
\mathbf{e}_{(3)}=\partial_r.
\end{equation}
Hence, $\mathbf{e}_{(3)}$ connects a point on a MOTS at a certain
instant of time with a corresponding point on the MOTS at the next
instant of time. Note that this choice of frame breaks down at the
poles of the sphere.  To apply formula \eqref{eq:compute_qij}, the
frame \eqref{eq:frame_H} on $H$ must be pushed forward to $\M$ by
means of the embedding $F$ in the standard way:
\begin{equation}
\mathbf{e}_{(3)}=(1,\partial_r F^1,\partial_r F^2,\partial_r F^3).
\end{equation}
This enables us to calculate $\mathbf{q}_{(i)(j)}$ using the 3-metric
on the Cauchy surface, and the lapse and shift.

Having calculated the matrix $\mathbf{q}_{(i)(j)}$ and assuming its
determinant to be positive, we can easily calculate the unit vector
$\hat{r}^a$.  It is simply the outward pointing unit spacelike vector
which is a linear combination of
$(\mathbf{e}_{(1)},\mathbf{e}_{(2)},\mathbf{e}_{(3)})$, and is
orthogonal to $\mathbf{e}_{(1)}$ and $\mathbf{e}_{(2)}$.  This
construction of $\hat{r}^a$ will also work in the timelike case, but
not in the null case where $\mathbf{q}_{(i)(j)}$ becomes degenerate.

\subsection{Angular momentum and mass}
\label{subsec:angmom}

Let $\fie^a$ be a rotational vector field on $H$ tangent to each
cross-section.\footnote{This means that $\fie^a$ is tangent to $S$,
  has closed integral curves, and is normalized so that its integral
  curves have an affine length of $2\pi$, and it vanishes at exactly
  two points on $S$.}  The angular momentum of a cross-section $S$
associated with $\fie^a$ is given by
\begin{equation}\label{eq:angmom}
J_S^{(\fie)} = -\frac{1}{8\pi }\oint_SK_{ab}\fie^a\hat{r}^b d^2V\,.
\end{equation}
We refer to \cite{Ashtekar03a} for a justification for this formula.
The interpretation of $J_S^{(\fie)}$ as angular momentum is most clear
cut when $\fie^a$ is a rotational symmetry on $H$, i.e., when
$\Lie_\fie K_{ab} = 0$ and $\Lie_\fie q_{ab} = 0$.  See
\cite{Dreyer02a} for a method of finding Killing vectors suitable for
numerical implementation.  Booth and Fairhurst have shown that this
formula also arises from a Hamiltonian calculation \cite{Booth05}. As
we shall see below, $J_S^{(\fie)}$ is also gauge invariant when
$\fie^a$ is only divergence free, and not necessarily a symmetry
vector.  However, $J_S^{(\fie)}$ may not be meaningful for more
general $\fie^a$.

If a cross-section $S$ has radius $R_S$ and angular momentum
$J_S^{(\fie)}$, we can meaningfully talk about the mass: 
\begin{equation}\label{eq:mass}
M_S^{(\fie)} = \frac{1}{2R_S}\sqrt{R_S^4 + 4(J_S^{(\fie)})^2} \,.
\end{equation}
This mass has the same dependence on the area and angular momentum as
in the Kerr solution.  There is a meaningful balance law for the mass
and furthermore, it satisfies a physical process version of the first
law \cite{Ashtekar02a,Ashtekar03a}.

Equation (\ref{eq:angmom}) uses the metric $q_{ab}$ and $K_{ab}$ and
extrinsic curvature of the dynamical horizon.  It is more convenient
to recast this in terms of the metric $\bar{q}_{ab}$ and extrinsic
curvature $\bar{K}_{ab}$ of the partial Cauchy surface $\Sigma$ (see
figure \ref{fig:cauchy}).  It is also more convenient to work with the
null
normals $(\bar{\ell}^a,\bar{n}^a)$ defined in equation
(\ref{eq:nullnormals}).  It is clear that $(\bar{\ell}^a,\bar{n}^a)$
must be related to the old null normals $(\ell^a,n^a)$ by a boost
transformation, i.e., there must exist a positive function $f$ on $S$
such that
\begin{equation}\ell^a = f\bar{\ell}^a \qquad \textrm{and} \qquad n^a =
f^{-1}\bar{n}^a \,. 
\end{equation}
After some simple algebra, equation (\ref{eq:angmom}) can be written
as:
\begin{equation}
J_S^{(\fie)} = -\frac{1}{8\pi} \left(\oint_S
  \bar{K}_{ab}R^a\fie^b\,d^2V + \oint_S
\Lie_\fie \ln f \, d^2V\right) \nonumber \,.
\end{equation}
The second integral vanishes precisely when $\fie^a$ is divergence
free, i.e., when $\fie^a$ is a symmetry of the area element on $S$.
In this case:
\begin{equation}\label{eq:finalangmom}
J_S^{(\fie)} = -\frac{1}{8\pi } \oint_S \bar{K}_{ab}R^a\fie^b\,d^2V \,.
\end{equation}
In particular, this will be true when $\fie^a$ is a symmetry of the
metric $\tilde{q}_{ab}$, but the divergence free condition is much
weaker than this.  For example, following \cite{Ashtekar:2004cn}, we
can always construct a divergence free vector field on a 2-sphere even
in the absence of axisymmetry as follows.  Let $h$ be any smooth
function on $S$, and $g$ another smooth function satisfying
$\tilde{\epsilon}^{ab}\partial_ah\partial_bg = 0$, where
$\tilde{\epsilon}_{ab}$ is the volume form on $S$.  It is easy to
check explicitly that the following vector field is automatically
divergence free:
\begin{equation}\label{eq:divfreephi}
\tilde{\fie}^a = g\tilde{\epsilon}^{ab}\partial_b h \,.
\end{equation}
The integral curves of $\tilde{\fie}^a$ are the level curves of $h$.
In particular, if $h$ is chosen to be a geometric quantity such as,
say, the curvature $\tilde{R}$, and $g$ chosen such that
$\tilde{\fie}^a$ has affine length $2\pi$, then $\tilde{\fie}^a$ will
coincide with an axial Killing vector, if it exists.  Therefore,
$\tilde{\fie}^a$ can be viewed as an \textit{ersatz} axial symmetry
vector field even in the absence of axisymmetry.

However, we haven't as yet satisfactorily implemented the above
construction due to numerical difficulties arising from errors in
taking derivatives of the scalar curvature.  Furthermore, the $\fie^a$
coming from eq.  (\ref{eq:divfreephi}) may not look like a rotational
vector field; in particular it may vanish at more than just two points
on the sphere even when $S$ is close to axisymmetry.\footnote{We thank
  Ivan Booth for this comment.} This is work in progress.  The results
presented below all use the method described in \cite{Dreyer02a} of
finding Killing vectors based on the Killing transport equations. This
reduces the problem of finding Killing vectors on a sphere to the
diagonalization of a $3\times 3$ matrix, and integrating a
1-dimensional ordinary differential equation.  We have found this
method to be quite reliable for the cases when the horizon is
sufficiently close to axisymmetry, even in cases when the coordinate
system is not adapted to the axial symmetry.  Thus, it works well for
the head-on collision and axisymmetric neutron star collapse, but only
at very early and late times for a non-axisymmetric black hole
collision.  This caveat only affects the example of section
\ref{subsec:puncture}. It is important to keep in mind that this
Killing transport method is not reliable for checking whether the
horizon is close to axisymmetry; this requires an independent
calculation of $\Lie_\fie \tilde{q}_{ab}$ to verify that it is
sufficiently small.  Finally, we emphasize that this method is also
not guaranteed to produce a divergence free rotational vector field;
this must also be checked independently.

\subsection{Multipole moments}
\label{subsec:multipoles}

The notion of multipole moments play a very important role in
Newtonian gravity and classical electrodynamics.  Let us focus on
classical electrodynamics in Minkowski space with axisymmetric charge
and current distributions ${\rho}$ and $j_a$ respectively, given on a
sphere $S$ of radius $R_S$.  Let $(\theta,\phi)$ be coordinates on
$S$; $\rho$ and $j_a$, being axisymmetric, are functions only of
$\theta$.  The electric multipoles $E_n$ and magnetic multipoles $B_n$
are respectively defined as
\begin{eqnarray}
E_n &=& R_S^n\oint \rho P_n(\cos\theta) d^2V \,,\\
B_n &=& -R_S^{n+1}\oint_S \left(\vec{j} \times \vec{\tilde{\partial}}
P_n(\cos\theta)\right) \cdot \hat{n} \,d^2V \,, \end{eqnarray}
where $P_n$ is the $n^\mathrm{th}$ Legendre polynomial,
$\tilde{\partial}$ denotes the standard derivative operator on a
sphere, and $\hat{n}$ is the unit outward normal to the sphere.  For
black holes, the analogs of the electric and magnetic multipole
moments are respectively the mass and angular momentum multipole
moments.  Motivated by this analogy, there exist meaningful
definitions of the source multipole moments for an isolated horizon
\cite{Ashtekar04a}.  Roughly speaking, these definitions correspond to
taking the moments of the free data on an axisymmetric isolated
horizon, and knowledge of these moments is sufficient to construct the
entire horizon geometry.

For dynamical horizons, we can generalize the construction of
\cite{Ashtekar04a} to construct a set of multipole moments which
capture the geometry of a dynamical horizon at any instant of time,
and which are furthermore equal to the isolated horizon multipole
moments when the black hole is isolated.  The analog of charge density
is (proportional to) the scalar curvature on $S$:
\begin{equation}
\rho_S = \frac{1}{8\pi}M_S\tilde{\R}\,,
\end{equation}
and the angular momentum current is 
\begin{equation}
j_a = -\frac{1}{8\pi }\tilde{q}_a^c\bar{K}_{cb}R^b \,.
\end{equation}
The moments of these quantities will give the desired multipole
moments.  We could also use
$\tilde{q}_a^cK_{cb}\hat{r}^b$ instead of
$\tilde{q}_a^c\bar{K}_{cb}R^b$ above; the two expressions are related
by a boost transformation.  Just as for angular momentum, the final
expressions for the multipole moments given below will be boost
invariant if the $\fie^a$ used in their definition is divergence free.   
To define the moments, we need a preferred coordinate system
on $S$ so that we can define the preferred spherical harmonics.  

The construction of the preferred coordinate system $(\theta,\phi)$ on
$S$ is the same as given in \cite{Ashtekar04a}: $\phi\in [0,2\pi)$ is
the affine parameter along $\fie^a$ and $\zeta:=\cos\theta \in [-1,1]$
is defined by the condition   
\begin{equation}\label{eq:zetadef} 
\tilde{D}_a\zeta = \frac{1}{R_S^2}\tilde{\epsilon}_{ba}\fie^a \,.
\end{equation}
The freedom to add a constant to $\zeta$ is removed by requiring its
integral over $S$ to vanish: $\oint_S \zeta \,d^2V = 0$.  When applied
to a Kerr black hole, these invariant coordinates turn out to be the
same as the usual Boyer-Lindquist $(\theta,\phi)$ coordinates. 

The mass and angular multipole moments are then respectively: 
\begin{eqnarray}
M_n &=& \frac{R_S^n M_S}{8\pi}\oint_S\left\{\tilde{\R}P_n(\zeta)\right\} d^2V \,,\\
J_n &=& -\frac{R_S^{n+1}}{8\pi} \oint_S
\left\{\tilde{\epsilon}^{ab}(\partial_b P_n(\zeta)) K_{ac}R^c
\right\}d^2V \nonumber \\
&=& \frac{R_S^{n-1}}{8\pi} \oint_S P_n^\prime(\zeta)\bar{K}_{ab}\fie^aR^b\,d^2V
\end{eqnarray}
where $P_n^\prime(\zeta) = dP_n(\zeta)/d\zeta$.  We have used equation
(\ref{eq:zetadef}) to obtain the final expression for $J_n$ above.
This form clarifies the relation of $J_n$ to the angular momentum and
also demonstrates the gauge invariance of $J_n$ when $\fie^a$ is
divergence free.  Using the Gauss-Bonnet theorem, it is trivial to
check that $M_0=M_S$ and $J_1=J_S$.  $J_0$ vanishes because we do not
consider any topological defects. Furthermore, these expressions are
well suited for numerical computation because they involve only
quantities on the Cauchy surface and an integral over the MOTS.

\subsection{The energy and angular momentum fluxes}
\label{subsec:energyflux}

Hawking's area theorem shows that if matter satisfies the null energy
condition, then the area of the event horizon can never decrease.
This is one of the central results of black hole physics, and it leads
to the classical picture of the black hole growing inexorably as it
swallows matter and radiation.  Therefore, one might expect there to
be a balance law relating the increase in area to fluxes of matter and
radiation crossing the event horizon.  However, the teleological
nature of event horizons is again a problem; there cannot exist any
such local balance law for the area of the event horizon.  A clear
example is seen in the Vaidya spacetime where the event horizon is
formed in flat space and its area increases in anticipation of matter
falling into the black hole at a later time; see
\cite{Ashtekar:2004cn} for a discussion.  

For DHs, it is possible to obtain an exact balance law for the area
increase \cite{Ashtekar02a,Ashtekar03a}; i.e., given two
cross-sections $S_1$ and $S_2$ with radii $R_1$ and $R_2$
respectively, and with $S_2$ lying to the outside of $S_1$, the
increase in the radius is given by the sum of the energy flux due to
matter ($\Fm$) and gravitational radiation ($\Fg$), both of which are
manifestly positive.:
\begin{equation}\label{eq:areabalance}
\frac{R_2 - R_1}{2} = \Fm + \Fg\,,
\end{equation}
where
\begin{eqnarray} 
\Fm &=& \int_{H}\sqrt{2} T_{ab}\hat{\tau}^a \ell^b dR\,d^2V \,,\\
\Fg &=& \frac{1}{8\pi }\int_{H} \left\{|\sigma_\ls|^2 +
|\zeta|^2\right\} dR\,d^2V\,.
\end{eqnarray}
Here $|\sigma_\ls|^2:=\sigma_{\ls ab}\sigma_\ls^{ab}$,
$|\zeta|^2:=\zeta_a\zeta^a$ where $\zeta^a$ is a vector on $S$ defined
as
\begin{equation}\zeta^a:=\sqrt{2}\tilde{q}^{ab}\hat{r}^c\nabla_c\ell_b \,, 
\end{equation}
and $d^2V$ is the natural geometric volume element on $H$.
The extra factors of 2 and $\sqrt{2}$ in the above equations as
compared to the corresponding equations in \cite{Ashtekar03a}, arise
because of our normalization convention $\ell\cdot n = -1$;
\cite{Ashtekar03a} uses $\ell\cdot n = -2$.     

See \cite{Ashtekar03a} for additional reasons why $\Fg$ has the right 
properties to be viewed as the flux of gravitational radiation.
Equation (\ref{eq:areabalance}) is an exact statement about black
holes in full non-linear general relativity, and it is the analog of the
Bondi mass balance law at null infinity.   

From a numerical point of view, $\Fg$ is inconvenient to
calculate, especially when the horizon is settling down and is close
to being null.  First of all, we have direct access only to the
fiducial null normals $(\bar{\ell}^a, \bar{n}^a)$ defined in eq.
(\ref{eq:nullnormals}) and not to $(\ell^a, n^a)$ themselves.  The two
sets of null normals are related to each other by a boost
transformation $\ell^a = f\bar{\ell}^a$, $n = f^{-1}\bar{n}^a$.  Under
this transformation, $\sigma_{\ell} = f\sigma_{\bar{\ell}}$.
Similarly, it is easy to show that
\begin{equation}
  \zeta^a = f^2\bar{\kappa}^a - \bar{\omega}^a\,,
\end{equation}
where 
\begin{equation}
  \bar{\kappa}^a = \tilde{q}^{ab}\bar{\ell}^c\grad_c\bar{\ell}_b  \quad \textrm{and} 
  \quad \bar{\omega}^a =
  \tilde{q}^{ab}\bar{n}^c\grad_c\bar{\ell}_b\,. 
\end{equation}
Here $\bar{\kappa}^a$ and $\bar{\omega}^a$ are tangent to the
cross-sections of the DH.  When the DH approaches equilibrium,
$f\rightarrow\infty$. However, the value of $\Fg$ itself remains
finite.  All fields with a bar remain finite even when the horizon
becomes null even though $f$ diverges While this is not a problem
analytically, this does cause numerical errors in the transition to
equilibrium when we multiply a very small quantity on the horizon with
a very large one. This is consistent with the results of
\cite{Booth04a} where it is found that $|\sigma_{(\bar{\ell})}|^2$ is
the most important when the horizon is close to equilibrium.  

Let $t$ be the time coordinate used to label the Cauchy surfaces.
Using this coordinate, we can identify the divergence of various terms
appearing in $\Fg$.  We start by rewriting $\Fg$ as:
\begin{equation}
\Fg = \frac{1}{8\pi }\int_{H}\left\{|\sigma_\ls|^2 +
|\zeta|^2\right\} \frac{dR}{dt}d^2V\,dt\,.  
\end{equation}
The integrand on the right hand side can be expanded as
\begin{eqnarray}
&& \left(|\sigma_\ls|^2 + |\zeta|^2\right) \dot{R} =
\nonumber \\
&& \dot{R}f^4 |\bar{\kappa}|^2 + \dot{R}f^2
(|\sigma_{(\bar{\ell})}|^2 - \bar{\omega}\cdot\bar{\kappa}) +
\dot{R}|\bar{\omega}|^2 \,.
\end{eqnarray}
Let us look at the various terms in this expression.  First,
$\bar{\omega}^a$ can be shown to be equal to the angular momentum
current; for an axial symmetry vector $\varphi^a$, the angular
momentum is simply the integral of $\varphi^a\bar{\omega}_a$ over the
cross section of the MTT.  Thus, $\bar{\omega}_a$ need not vanish even
when the MTT becomes an isolated horizon.  The $|\bar{\omega}|^2$ term
in the flux can, in some sense, be viewed as the flux of rotational
energy entering the horizon.  Now consider $\bar{\kappa}^a$.  For an
isolated horizon, $\bar{\ell}^b\grad_b\bar{\ell}^a \propto
\bar{\ell}^a$ because in this case $\bar{\ell}^a$ is guaranteed to be
geodetic.  This implies $\bar{\kappa}^a = 0$.  On the dynamical
horizon side, we can choose suitable extensions of $\bar{\ell}^a$ (and
$\bar{n}^a$) away from the MTT so that $\bar{\kappa}^a=0$.  The shear
$\sigma_{(\bar{\ell})}$ on the other hand contains most of the
non-trivial information about the radiation falling into the black
hole.  It vanishes on an isolated horizon as it should, and it is
independent of any extensions of $\bar{\ell}^a, \bar{n}^a$ away from
the MTT.  Therefore, in the examples of section \ref{sec:example}, we
shall usually plot $\sigma_{(\bar{\ell})}$ to show the energy flux
falling into the horizon.  

The angular momentum also obeys a balance law similar to equation
(\ref{eq:areabalance}):
\begin{equation}
J_2 - J_1 = \Jm_{\fie} + \Jg_\fie
\end{equation}
where
\begin{eqnarray}
\Jm_{\fie} &=& -\int_{\Delta H} T_{ab}\hat{\tau}^a\fie^b
d^3V \,,\\
\Jg_{\fie} &=& -\frac{1}{16\pi }\int_{\Delta H}
P^{ab}\Lie_\fie q_{ab} d^3V 
\end{eqnarray}
where $P^{ab} :=K^{ab}-Kq^{ab}$.  Unlike the energy flux $\Fg$, the
angular momentum flux $\Jg$ is not positive definite.  Also, $\Jg$
vanishes when $\fie^a$ is an axial Killing vector on $H$.  Thus,
angular momentum is conserved in the axisymmetric vacuum case, as it
should be.

\section{Example numerical simulations}
\label{sec:example}

In this section, we apply the ideas discussed in the previous sections
to three concrete numerical simulations: i) A head-on collision of two
black holes starting with Brill-Lindquist initial data;  ii) A
non-axisymmetric black hole collision using puncture initial data with
non-vanishing linear momentum and iii) Axisymmetric collapse of a
neutron star.  Each of these three cases is quite well known in the
numerical relativity literature, and all have been well studied.
This section aims to further explore these examples using the tools
described in Section \ref{sec:numerics}.

\subsection{Head-on collision with Brill-Lindquist data}
\label{subsec:headon}

The Brill-Lindquist initial data \cite{Brill63} for binary black holes
represent initial data for two non-spinning black holes without any
orbital angular momentum.  The reader can consult a review on initial
data, such as \cite{Cook00a1} for details.  Here we simply note that
these initial data are conformally flat and time-symmetric:
\begin{equation}
\hphy_{ab} = \psi^4\delta_{ab}\,,\qquad \Kphy_{ab} = 0 \,. 
\end{equation}
The manifold $\Sigma$ is $\mathbb{R}^3$ with two points removed (the
punctures).  The only equation to be solved is the flat space Laplace
equation for the conformal factor:
\begin{equation}
\Delta \psi = 0\,.
\end{equation}
Let $d$ denote the shortest distance between the two punctures as
measured with respect to the fictitious flat background metric
$\delta_{ab}$; the physical proper distance between the punctures is
actually infinite.  It was shown in \cite{Brill63} that each of the
punctures is actually an asymptotically flat region.  The total ADM
mass of the common asymptotic region is
\begin{equation}
m_\ADM = 2\alpha_{(1)} + 2\alpha_{(2)}\,, 
\end{equation}
and the ADM masses of the two punctures are
\begin{eqnarray}
m_{(1)}^\ADM &=& 2\alpha_{(1)} + \frac{2\alpha_{(1)}\alpha_{(2)}}{d}
\,\\
m_{(2)}^\ADM &=& 2\alpha_{(2)} + \frac{2\alpha_{(1)}\alpha_{(2)}}{d}\,.\\
\end{eqnarray}
These are exact results, irrespective of the distance $d$ between the
punctures.  In the next two sub-sections, we look at two different
regimes (i) the far limit when $d$ is large and (ii) the merger of the
two holes starting from relatively small values of $d$.

\subsubsection{The far limit}
\label{subsubsec:farlimit}

Before presenting the results from the numerical evolution of this
data, it is instructive to look at a special case which is amenable to
analytic treatment, namely, in the far limit where the separation
between the holes is very large: $d \gg \alpha_{(1)}, \alpha_{(2)}$.
In this case, there are two MOTSs surrounding each of the punctures
without any common MOTS surrounding them.  The angular momenta of the
two black holes are trivially zero because the extrinsic curvature 
vanishes.  What about the mass?  Should $m_{(1)}^\ADM$ and
$m_{(2)}^\ADM$  be identified with the masses of the black holes?
There are three difficulties with this.  First, these ADM masses also
include contributions from radiation present in the respective
asymptotic regions. 
Secondly, if this identification is correct, $m_{(i)}^\ADM$
($i=1,2$) is supposed to be the mass of the black hole 
for all values of $d$, even when the two black holes are
very close to each other.  Shouldn't the mass of the black holes in
this regime also include, say, contributions from the tidal
distortions produced by the other hole?  Finally, the strategy of
using the asymptotic regions to define black hole masses is not
applicable generally, say in the case when there are matter fields and
the topology of $\Sigma$ is just $\mathbb{R}^3$, or in Misner data
\cite{Misner63} where the two black holes do not have their own
individual asymptotic regions. 

From the isolated/dynamical horizon perspective, since the black holes
have zero angular momentum, from equation (\ref{eq:finalangmom}), the
irreducible mass is the correct measure of mass in this case: $m_{(i)}
= \sqrt{a_{(i)}/16\pi}$ 
where $a_{(i)}$ is the area of the MOTS around each of the punctures.
Let us then calculate the mass of the black holes as a power series in 
$1/d$.  To simplify calculations, put the origin of coordinates at the
location of the first puncture and the other puncture on the $z$-axis
at $(0,0,d)$.   Introduce the usual spherical coordinates
$(r,\theta,\phi)$ so that the conformal factor becomes explicitly
\begin{equation}
\phi(r,\theta) = 1 + \frac{\alpha_{(1)}}{r} +
\frac{\alpha_{(2)}}{r}\left( 1 - \frac{2d\cos\theta}{r} +
\frac{d^2}{r^2} \right)^{-\frac{1}{2}}\,.
\end{equation}
We see that due to axisymmetry, there is no dependence on $\phi$.  Let
the surface of the FMOTS around the origin be given by the equation
$r=h(\theta)$. In the limit when $d\rightarrow \infty$, the initial
data reduces to Schwarzschild in isotropic coordinates so that the
horizon is located at $r=\alpha_{(1)}$.  

Higher order effects can also be explicitly calculated. It turns out
\cite{Krishnan02} that up to $\mathcal{O}(d^{-3})$, the location of
the MOTS is given by
\begin{eqnarray}
r &=& \alpha_{(1)} -\frac{\alpha_{(1)}\alpha_{(2)}}{d} +
\frac{\alpha_{(1)}\alpha_{(2)}}{d} (\alpha_{(2)} -
\alpha_{(1)}\cos\theta)  \nonumber \\
&-& \frac{\alpha_{(1)}\alpha_{(2)}}{3} \left(\alpha_{(2)}^2 -
3\alpha_{(1)}\alpha_{(2)}\cos\theta \right. \nonumber \\ & & +
\left.\frac{5}{7}\alpha_{(1)}^2P_2(\cos\theta)\right) + \mathcal{O}(d^{-4})
\end{eqnarray}
where $P_2$ is the second Legendre polynomial.  Using this result, the
horizon mass $m_{(i)} = \sqrt{a_{(i)}/16\pi}$ can be calculated and,
somewhat surprisingly, the mass is the same as the ADM mass even up to
third order: 
\begin{equation}
m_{(1)} = 2\alpha_{(1)} + \frac{2\alpha_{(1)}\alpha_{(2)}}{d} +
\mathcal{O}(d^{-4})\,. 
\end{equation}
This relation was verified numerically for a sequence of BL data with
different values of $d$.  However, we did not have sufficient
resolution to estimate the leading order deviation between $m_{(1)}$
and $m_{(1)}^\ADM$.  Similarly, the shear of the horizon vanishes up to
third order indicating that the individual horizons are isolated to an
excellent approximation.  As we shall see below, the individual
horizons are isolated even for relatively small values of $d$ once the
common MOTS has formed.

\subsubsection{Numerical results for the merger phase}
\label{subsec:resultsbl}

We performed a numerical evolution starting with Brill-Lindquist
initial data. Working in units where the total ADM mass is unity, the
punctures were located at $z=\pm 0.5$, and the individual black holes
had equal masses.  Thus $2\alpha_{(1)} = 2\alpha_{(2)}=0.5$.  The
domain had an explicit octant symmetry and extended up to $x,y,z=96$.
Near the outer boundary the spatial resolution was $h=1.6$, and near
the punctures we used mesh refinement to increase the resolution
successively up to $h=0.0125$, so that the individual horizon
diameters contained initially $32$ grid points.  We used fourth order
accurate spatial differencing operators, and a third order
Runge--Kutta time integrator.

We excised \cite{Alcubierre00a} coordinate spheres with a radius of
$r_e=0.0625$ about the punctures from the domain, corresponding to a
diameter of $10$ grid points.  We used the AEI BSSN formulation
\cite{Alcubierre00a,Alcubierre01a} for time evolution, using the
boundary conditions also described in \cite{Alcubierre00a}.  These
boundary conditions are known to be incompatible with the Einstein
equations.
We used a $1+\log$
slicing condition \cite{Alcubierre02a} starting from $\alpha=1$, and a
zero shift.  This makes both the individual and the outer common
horizon grow in coordinate space.  We used the Cactus framework
\cite{Goodale02a, cactusweb1}, the Carpet mesh refinement driver
\cite{Schnetter-etal-03b, carpetweb}, and the \texttt{CactusEinstein}
infrastructure.  We located the apparent horizon surfaces with
J. Thornburg's \texttt{AHFinderDirect}
\cite{Thornburg2003:AH-finding,Thornburg95}.

\begin{figure} 
  \includegraphics[width=0.45\textwidth]{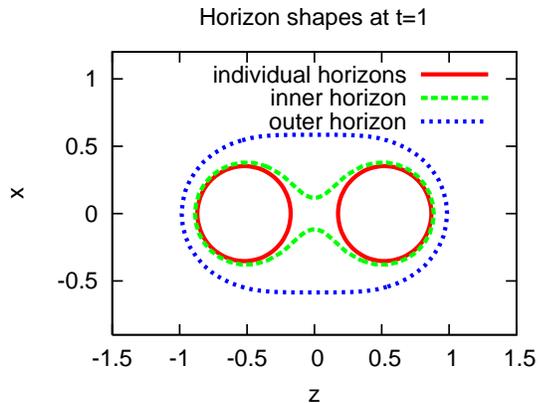}
  \caption{Coordinate shapes of the horizons at $t=1$ in the $xz$
    plane.  A common horizon has formed, and the inner and outer
    common horizons have already separated.  Compare figure
    \ref{fig:multibh}.}
  \label{fig:headon-shape-t1}
\end{figure}
\begin{figure} 
  \subfigure[$t=0.6$]
  {\includegraphics[width=0.45\textwidth]{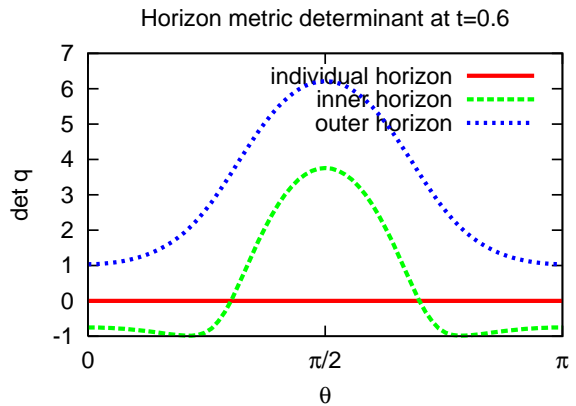}}
  \subfigure[$t=1$]
  {\includegraphics[width=0.45\textwidth]{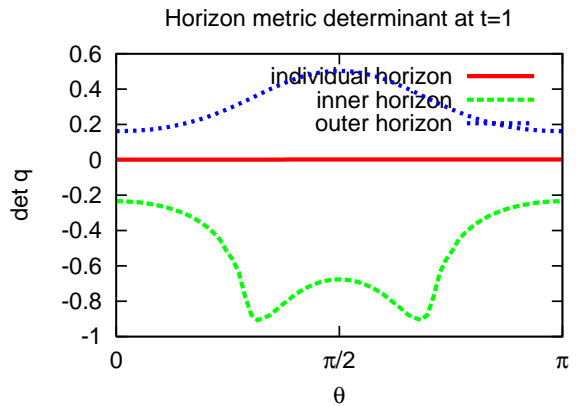}}
  \caption{Determinant of the horizon world tube's three-metric vs.\
    latitude $\theta$ at $t=0.6$ and $t=1$.  The individual MTTs are
    null, i.e., $\mathrm{det}\; \tilde{q} =0$ (up to numerical errors).
    The common outer MTT is spacelike (i.e.,
    $\mathrm{det}\; \tilde{q} > 0$) and it tends to null at late times.
    The inner common MTT is partially timelike at $t=0.6$; later it
    becomes completely timelike.}
  \label{fig:headon-3det}
\end{figure}
In this setup, the apparent horizon has two disconnected components in
the initial data, and a common MOTS forms shortly after $t=0.5$. The individual
horizons are null up to numerical errors (consistent with the result
on the smallness of $\sigma_\ls$ in the far limit), and their masses
are essentially constant up to numerical error.  As
discussed in section \ref{subsubsec:signature1} and figure
\ref{fig:multibh}, the common MOTS forms initially as a single surface
but then bifurcates: an outer horizon
which is strictly-stably-outermost, and an inner one which becomes
strictly untrapped on being deformed inwards.  Figure
\ref{fig:headon-shape-t1} shows the shapes of the individual and the
inner and outer common MOTSs at time $t=1$, where the inner and outer
common MTTs have already noticeably separated. As expected, the outer MTT is purely spacelike while the inner
MTT, being spacelike initially, becomes partly timelike quickly.
Figure \ref{fig:headon-3det} shows the  
horizon world tube metric signature at $t=0.6$ and $t=1$.  At later
times, the outer MTT tends to become null (as expected), while the
inner MTT becomes completely timelike, and then becomes so distorted
at about $t=1.2$ that it cannot be reliably tracked any more.  This
coordinate distortion is already evident in figure
\ref{fig:headon-shape-t1}, and the horizon discretisation used in the
apparent horizon finder is inaccurate near the neck of the inner
horizon \cite{Thornburg2003:AH-finding}.
\begin{figure} 
  \includegraphics[width=0.45\textwidth]{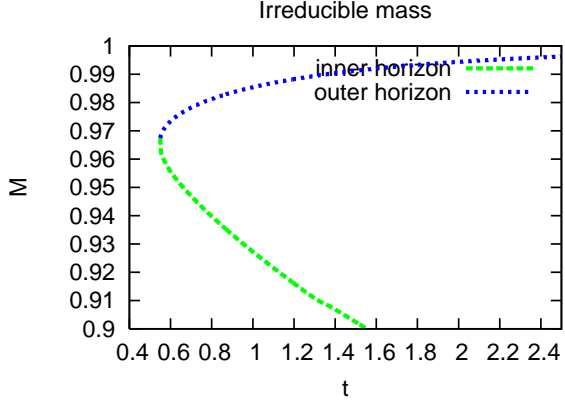}
  \caption{Irreducible mass vs.\ time for the individual and the
    common MTTs. The outer common MTT grows and accretes mass,
    while the inner MTT shrinks and loses mass.}
  \label{fig:headon-M}
\end{figure}
Figure \ref{fig:headon-M} shows the time evolution of the masses $M =
\sqrt{A_S/16\pi}$ of the individual and the common horizons (in this
case, the angular momentum vanishes identically). If $M^\infty$ is the
asymptotic value of the mass of the outermost horizon at late times,
then $M_\ADM - M^\infty$ is, in principle, a reliable way of
estimating the amount of energy radiated away to infinity in the form
of gravitational waves.  This difference could be used as a
consistency check on other estimates using the extracted waveforms at
large distances from the black holes. However, our emphasis in this
paper is on the dynamics of the merger and not on long duration stable
evolutions. Our simulations do not last long enough to estimate
$M^\infty$ reliably.

Another feature of the horizons, shown in figure \ref{fig:headon-M},
is that while the common outer MTT increases in area as expected, the
area of the common inner MTT decreases monotonically.  This is
explained as follows.  Initially, when the common MOTS is just formed,
by continuity with the outer MTT, the inner MTT is spacelike for a
very short duration (much before $t=0.6$) and it is thus a DH for this
duration.  However, this DH is being traversed in the \emph{inwards}
direction (i.e., along $-\hat{r}^a$) so that its area appears to
decrease.  Shortly after its formation, the inner MTT becomes partly
timelike and later fully timelike.  Recall that for a TLM, the area
decreases if $\Theta_\ns< 0$.  Thus, both the spacelike and timelike
portions of the inner MTT contribute to its monotonic area decrease.
This behavior of the outer MTT is roughly similar to what was found in
\cite{Booth05a} for spherically symmetric horizons; however due to
spherical symmetry, the horizons in \cite{Booth05a} did not have any
cross sections of mixed signature.

\begin{figure} 
  \subfigure
  {\includegraphics[width=0.45\textwidth]{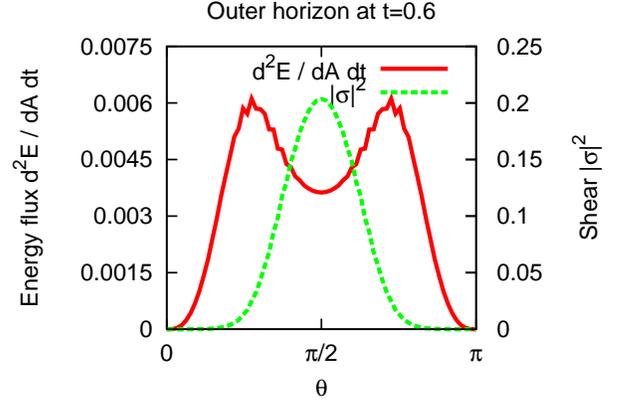}}
  \subfigure
  {\includegraphics[width=0.45\textwidth]{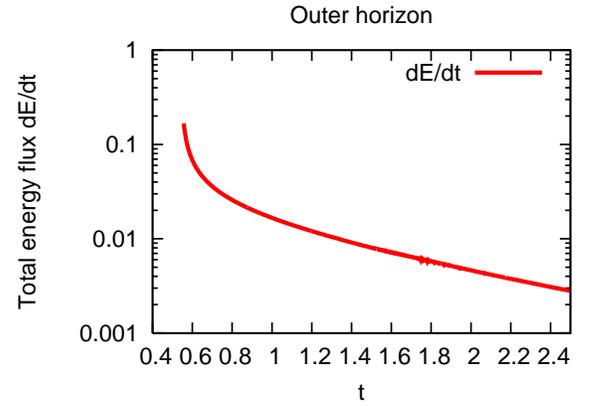}}
  \caption{Energy flux and shear $|\sigma_{(\bar{\ell})}|^2$ through
    the outer common horizon vs.\ latitude $\theta$ at $t=0.6$, and
    the total energy flux vs.\ time.  The shear vanishes at the poles
    and the black hole settles down exponentially.}
  \label{fig:headon-energy-flux}
\end{figure}
Figure \ref{fig:headon-energy-flux} demonstrates how the common outer
apparent horizon grows. The energy flux vanishes at the poles, and the
shear (but not the total flux) is maximum at the equator.  The horizon
is spacelike all the time, but it becomes increasingly isolated at
late times as it approaches equilibrium.  Thus the rate of area
increase becomes smaller and the fluxes also becomes correspondingly
smaller.

\begin{figure} 
  \subfigure
  {\includegraphics[width=0.45\textwidth]{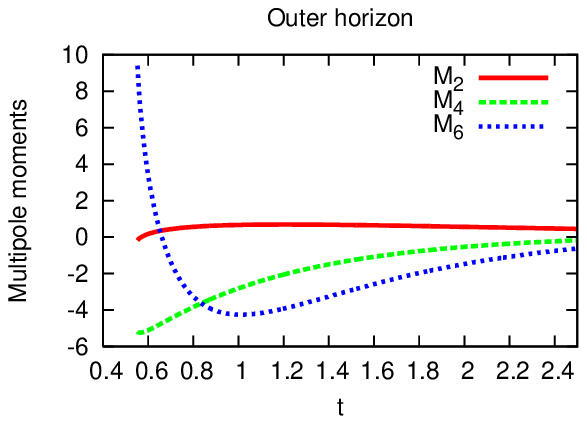}}
  \subfigure
  {\includegraphics[width=0.45\textwidth]{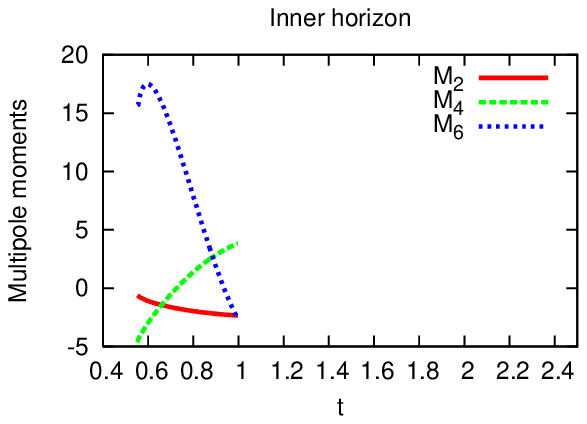}}
  \caption{Some mass multipole moments vs.\ time for the inner and
    outer MTTs for the head-on collision. The multipole moments for
    the outer horizon all approach their Schwarzschild values (i.e., 0)
    but the inner horizon does not seem to do so.}
  \label{fig:headon-multipole}
\end{figure}

Let us now consider the higher mass multipoles $M_n$ (all the $J_n$s
vanish identically).  Here, since all quantities are symmetric with
respect to a reflection about the equatorial plane, $M_n=0$ for odd
$n$.  Figure \ref{fig:headon-multipole} plots the mass quadrupole
moment $M_2$ and also $M_4$ and $M_6$ of the outer and inner common
MTTs as a function of time.  We expect that the black hole should
eventually settle down to a Schwarzschild solution by radiating away
all of its higher multipole moments.  Clearly, for the outer MTT,
$M_2, M_4$ and $M_6$ all become smaller with time, approaching zero.
However, the run did not last long enough for us to obtain the
asymptotic fall-off rate.  It is interesting to note that, as far as
we can tell, the multipole moments for the inner MTT do \emph{not}
vanish asymptotically.  This tells us that the spacetime near the
inner MTT is not close to Schwarzschild even at late times.  At even
later times, all the inner horizons presumably cease to exist (see
next paragraph) and the spacetime approaches Schwarzschild everywhere.

We conclude this section with some remarks on the eventual fate of the
inner MTT.  First of all, as expected, the outer MTT eventually
settles down and approaches future timelike infinity.  The inner MTT
shrinks and approaches the two individual horizons which are
essentially stationary.  It is interesting to speculate on how, if at
all, the inner MTT will merge with the two individual MTTs.  Does the
inner MTT ``pinch off'' into two individual horizons?  If the inner
MTT is indeed the one predicted by \cite{Schoen04}, then it has
\textit{a priori} curvature bounds.  If these curvature bounds are
maintained in the limit, then the inner horizon cannot pinch off.  It
is more likely that the two individual MTTs merge first with each
other and then later, perhaps also with the inner MTT.  It would be
interesting to investigate this question further.  If the inner MTT
does indeed merge smoothly with the two individual MTTs, then the set
of all MTTs in this case would form one single smooth 3-manifold.
Furthermore, the area of the cross-section of this manifold would be
monotonic in the \emph{outward} direction -- traversing this manifold
in the outward direction means going forward in time on the individual
and outer MTTs, and backward in time on the inner MTT.

We are not able to settle these issues numerically in a conclusive
manner because the inner MTT becomes so distorted at late times that
the AH tracker is no longer able to track it.  This is because the AH
tracker can only locate star-shaped surfaces and, as is clear from
figure \ref{fig:headon-shape-t1}, the inner MTT will not necessarily
be star-shaped at later times. Furthermore, our gauge choice in which
we allow the outer MTT to grow in coordinate space, makes the inner
MTT shrink and therefore harder to resolve at later times.

\subsection{Non-axisymmetric black hole collision}
\label{subsec:puncture}

The head-on collision described above does not incorporate any effects
of angular momentum.  In this section, we remove the restriction of
axisymmetry by taking initial configurations in which the black holes
are orbiting around each other.  We use the so called ``puncture''
data introduced by Brandt and Brügmann \cite{Brandt97b}, which is a
generalization of the Brill-Lindquist construction.  The data is still
taken to be conformally flat, but now no longer assumed to be time
symmetric.

We performed a numerical evolution of puncture initial data
corresponding to the innermost stable circular orbit as predicted in
\cite{Cook94}.  This model was also studied as ``QC-0'' with the
Lazarus perturbative matching technique \cite{Baker:2002qf,
  Baker:2003ds} and later in
\cite{Alcubierre2003:pre-ISCO-coalescence-times, Campanelli:2005dd,
  Baker05a, Campanelli:2006gf, Herrmann:2006ks}.  In our setup, the
punctures were located at $x=\pm 1.168642873$, and their mass
parameters were $m=0.453$, and their momenta were $p_y=\pm
0.3331917498$.  The domain had an explicit rotating quadrant symmetry
and extended up to $x,y,z=10$.  Near the outer boundary the spatial
resolution was $h=0.4$, and near the punctures we used mesh refinement
to increase the resolution successively up to $h=0.025$, so that the
individual horizon diameters contained initially $16$ grid points.  We
used fourth order accurate spatial differencing operators, and a third
order Runge--Kutta time integrator.

We excised \cite{Alcubierre00a} coordinate spheres with a radius of
$r_e=0.075$ about the punctures from the domain, corresponding to a
diameter of $6$ grid points.  We used again the AEI BSSN formulation
\cite{Alcubierre00a, Alcubierre01a} for time evolution, a $1+\log$
slicing condition \cite{Alcubierre02a} starting from a lapse that is
one at infinity and zero at the punctures, and a $\Gamma$ driver shift
condition, starting from a rigid co-rotation with an angular velocity
of $\omega=0.06$.  We also used a drift correcting shift term similar
to \cite{Bruegmann:2003aw, Alcubierre2003:co-rotating-shift} to keep
the individual horizons centered about their initial locations.

As previously, we used the Cactus framework \cite{Goodale02a,
  cactusweb1}, the Carpet mesh refinement driver
\cite{Schnetter-etal-03b, carpetweb}, and the \texttt{CactusEinstein}
infrastructure.  We solved the initial data equation with M. Ansorg's
\texttt{TwoPuncture} solver \cite{Ansorg:2004ds}, and we located the
apparent horizon surfaces with J. Thornburg's \texttt{AHFinderDirect}
\cite{Thornburg2003:AH-finding}.

This setup contains two initially separated horizons that rotate
around each other for a fraction of an orbit before a common horizon
forms \cite{Baker:2002qf, Alcubierre2003:pre-ISCO-coalescence-times}.
Its ADM mass is $M_\ADM = 1.00788$, the initial proper horizon separation is
$L \approx 4.99\,M_\ADM$, and the horizons have initially the angular
momentum $J \approx 0.78\,M_\ADM^2$ and angular velocity $\Omega \approx
0.17/M_\ADM$.  The common apparent horizon forms at about $t=17.5$, which
we verified through pretracking \cite{Schnetter04}.

\begin{figure} 
  \includegraphics[width=0.45\textwidth]{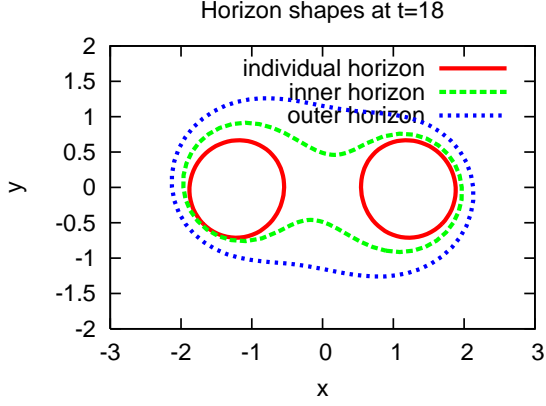}
  \caption{Coordinate shapes of the MOTSs at $t=18$ for the
    non-axisymmetric black hole collision.  Note that the individual
    horizons are locked in place through the co-rotating coordinate
    system and through an adaptive shift condition.}
  \label{fig:nonheadon-shape-t18}
\end{figure}
\begin{figure} 
  \includegraphics[width=0.45\textwidth]{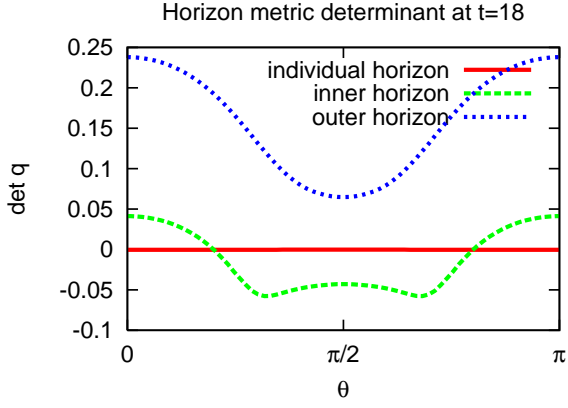}
  \caption{Determinant of the MTT three-metric at $t=18$. As in the
    head-on case, the outer MTT is purely spacelike while the inner
    MTT is partly spacelike and partly timelike. At later times, it
    becomes purely timelike. The individual MTTs are null at this
    time. }
  \label{fig:nonheadon-3det}
\end{figure}
\begin{figure} 
  \includegraphics[width=0.45\textwidth]{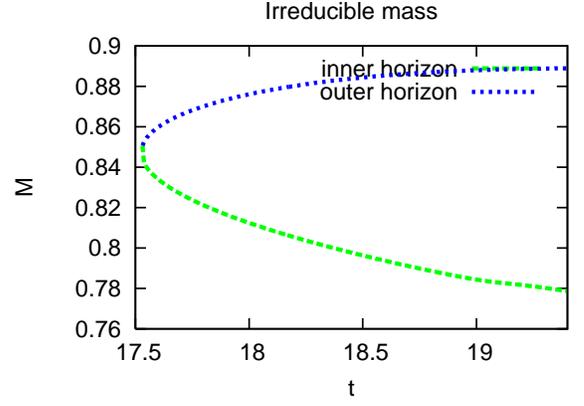}
  \caption{A plot of the irreducible mass $M_{irr} = \sqrt{A/16\pi}$
    as a function of time for the outer an inner MTTs in the
    non-axisymmetric black hole collision. As expected, the outer MTT
    has increasing area while the inner MTT shrinks.
  }\label{fig:nonheadon-irr}
\end{figure}

Figure \ref{fig:nonheadon-shape-t18} shows the shape of the various
MOTSs at a time $t=18$, a short while after the common horizon has
formed.  The qualitative behavior of the various MTTs is exactly the
same as in section \ref{subsec:resultsbl}.  Figure
\ref{fig:nonheadon-irr} shows the irreducible mass of the outer and
inner MTTs as a function of time.  Again, the behavior is
qualitatively the same as we saw in the head-on collision.

\begin{figure} 
  \vspace*{-0.05\textwidth}\hspace*{-0.05\textwidth}\includegraphics[width=0.55\textwidth]{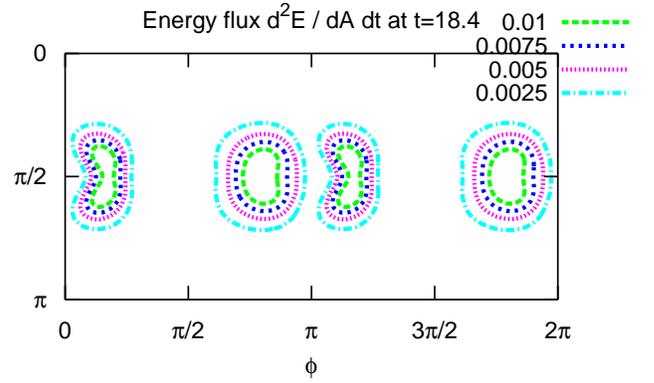}\vspace*{-0.05\textwidth}
  \vspace*{-0.05\textwidth}\hspace*{-0.05\textwidth}\includegraphics[width=0.55\textwidth]{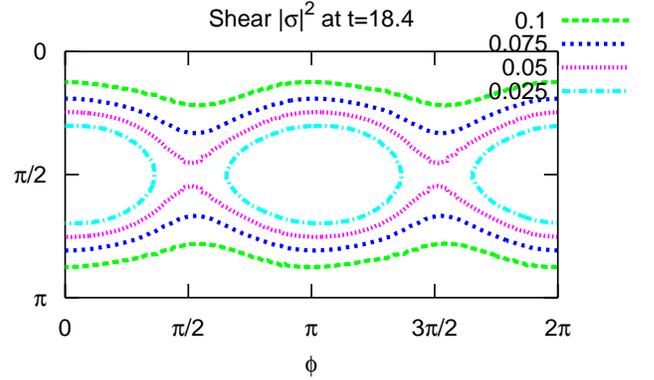}
  \caption{Energy flux through the common horizon and shear
    $|\sigma_{\bar{(\ell)}}|^2$ on the horizon at $t=18.4$ in
    $(\theta,\phi)$-coordinates.  The spacetime is not axisymmetric,
    and since it contained two inspiralling black holes, there is no
    mirror symmetry across the $x-z$ or $y-z$ planes either.  The
    energy flux shows this asymmetry clearly.}
  \label{fig:nonheadon-energyflux}
\end{figure}
Figure \ref{fig:nonheadon-energyflux} shows the flux of gravitational
wave energy
falling into the outer horizon at $t=18.4$ and also the shear
$|\sigma_{(\bar{\ell})}|^2$ at the same time, for the outer and
individual horizons. The 2-d contour
plots of the shear 
$|\sigma_{(\bar{\ell})}|^2$ and the total flux on the horizon shows in
detail how gravitational radiation is falling into the horizon.
Unlike in the head-on case (fig. \ref{fig:headon-energy-flux}), the
shear and the flux are now no longer axisymmetric.  Therefore, the
flux is no longer constant along the $\phi$ direction but its maxima
still lie on the equator.  The shear on the other hand, now has its
maximum on the poles and its minima lie on the equator. It would be
interesting to further investigate the behavior of
$|\sigma_{(\bar{\ell})}|^2$ and the energy flux as a function of time
and for different 
physical situations to gain a better understanding of how a black hole
grows.

\begin{figure} 
  \includegraphics[width=0.45\textwidth]{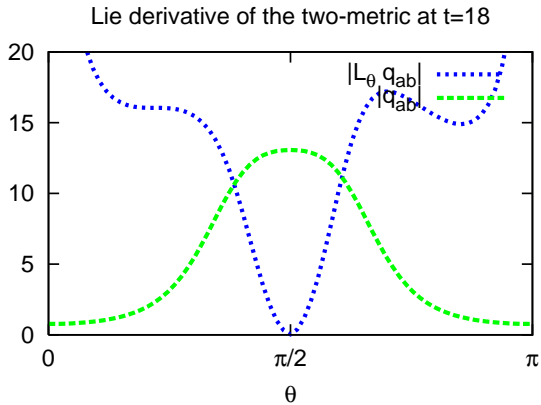}
  \caption{Lie derivative of the two-metric $\Lie_\fie \tilde{q}_{ab}$
    at $t=18$ on the $\phi=0$ line.  The two-metric $\tilde{q}_{ab}$
    is also shown for comparison.  The quantity shown in the plots are
    actually the norms $\sqrt{ \sum_{ab} (\Lie_\fie\tilde{q}_{ab})^2
    }$ and $\sqrt{ \sum_{ab} (\tilde{q}_{ab})^2 }$ in the coordinate
    system $(\theta,\phi)$ on the horizon.  The vector field $\fie$ is
    Killing on the equator (see main text), but not everywhere.  This
    shows that the horizon is not (yet) axisymmetric.  We expect it to
    become axisymmetric at later times.  Note that we have only shown
    the plots along the $\phi=0$ curve and we do not have axisymmetry
    here. }
  \label{fig:nonheadon-lq}
\end{figure}
Let us now turn to the rotational vector $\fie^a$ on the outer horizon
and the quantities
such as angular momentum, mass, and multipole moments associated with
it.  The simulation presented here was run only up to $t \approx
19.4$, and the final black hole has not settled down sufficiently, and
has not attained axisymmetry at this point.  Figure
\ref{fig:nonheadon-lq} shows the Lie derivative of the 2-metric
$\Lie_\fie \tilde{q}_{ab}$ on the horizon at $t=18$, where $\fie^a$ is
the Killing vector candidate found by the algorithm presented in
\cite{Dreyer02a}.  It is clear that
$\Lie_\fie\tilde{q}_{ab}$ is very far from 0 at this time.  This means
that the angular momentum, mass, and multipole moments associated with
this $\fie^a$ are not meaningful at this point.  This is to be
expected, since the final black hole should attain axisymmetry only on
a time scale set by the quasi-normal mode ringdown, which has a period
of $15.9 M_\ADM$ in this case.  It is interesting to see that our
Killing vector field candidate is indeed Killing on the equator.  This
is by construction, since we choose the Killing vector field candidate
by an integral along the equator; see
\cite{Dreyer02a}.  However, the vector field
$\fie^a$ is far from Killing away from the equator.

A word of caution is due here regarding the Killing vector finding
algorithm of \cite{Dreyer02a}.  First of all,
the algorithm only produces a candidate for a Killing vector, and an
independent check is required to see whether $\Lie_\fie\tilde{q}_{ab}$
is sufficiently small or not.  Furthermore, as mentioned previously,
this method reduces the problem of finding a Killing vector on a
sphere to diagonalizing a $3\times 3$ matrix followed by integrating a
1-dimensional ODE.  In particular, the method requires that one of the
eigenvalues of this matrix is sufficiently close to unity.  While this
is fine when the horizon is exactly axisymmetric, the subtlety arises
when the horizon is only approximately axisymmetric.  It is not clear
how close the eigenvalue must be to unity for the horizon to be
regarded as approximately axisymmetric.  Work is in progress to
understand this better and to also investigate an alternate method of
finding an appropriate $\fie^a$ as discussed in section
\ref{subsec:angmom}, which is guaranteed to produce a divergence free
vector.

\subsection{Axisymmetric gravitational collapse}
\label{subsec:nscollapse}

\subsubsection{The initial configuration}
\label{subsubsec:nsinitial}

Up to now, all of our examples have involved only vacuum spacetimes.
In this section, we present an example of the gravitational collapse
of a neutron star to form a black hole in an axisymmetric spacetime.
These simulations were performed using the \texttt{Whisky} code which
deals with the matter terms of the Einstein equations in the framework
of the \texttt{Cactus} toolkit.  Thus, the \texttt{Whisky} code solves
the conservation equations for the stress energy tensor $T_{ab}$ and
for the matter current density $J^a$:
\begin{equation}\grad^aT_{ab} = 0\,,\qquad \grad_aJ^a = 0\,. \end{equation}
For details about the \texttt{Whisky} code and the implementation of
the above equations, we refer the reader to \cite{Baiotti04} and
references therein.  Here we shall restrict ourselves to describing
the initial stellar configuration which is one of the configurations
studied in \cite{Baiotti04}.   

The neutron star is modeled as a uniformly rotating ball of perfect
fluid.  The equation of state is taken to be a $K=100$, $\Gamma=2$
polytrope so that the pressure $p$ and rest-mass density $\rho$ are
related according to $p = K\rho^\Gamma$.  The equilibrium
configuration is determined by the mass $M_\NS$, central density
$\rho_c$, and the angular momentum $J_\NS$; when necessary, the
subscript ${}^\NS$ is used in order to avoid any confusion with
previously defined symbols.  The model we take is the one denoted as
``D4'' in \cite{Baiotti04} which has $M_\NS = 1.86M_\odot$, $\rho_c =
1.934 \times 10^{15}\,\mathrm{g}\,\mathrm{cm}^{-3}$,
and $J_\NS=0.543M_\NS^2$.
This leads to a ratio of polar to equatorial coordinate radii of
$0.65$, a circumferential equatorial radius of $14.22$ km, and a
rotational frequency of $1295.34 \textrm{Hz}$.  This equilibrium
configuration turns out to be dynamically unstable. In practice, the
instability is induced by uniformly reducing the pressure slightly
throughout the star.

\subsubsection{Numerical results}
\label{subsec:nsresults}

We simulated the above system on a grid with an explicit rotating
octant symmetry.  The outer boundary was at $x,y,z=150$, and the grid
spacing near the outer boundary was $h=3$.  We used mesh refinement to
increase our spatial resolution in the center of the domain to
$h=0.375$ at the initial time, and progressively introduced more mesh
refinement levels to increase the central resolution up to
$h=0.046875$ as the neutron star collapsed, based on the maximum
density in the star \cite{Baiotti04b, Ott06a}.  We also apply third
order Kreiss--Oliger dissipation \cite{Kreiss73} to the spacetime (but
not the hydrodynamics) variables.

\begin{figure} 
  \includegraphics[width=0.45\textwidth]{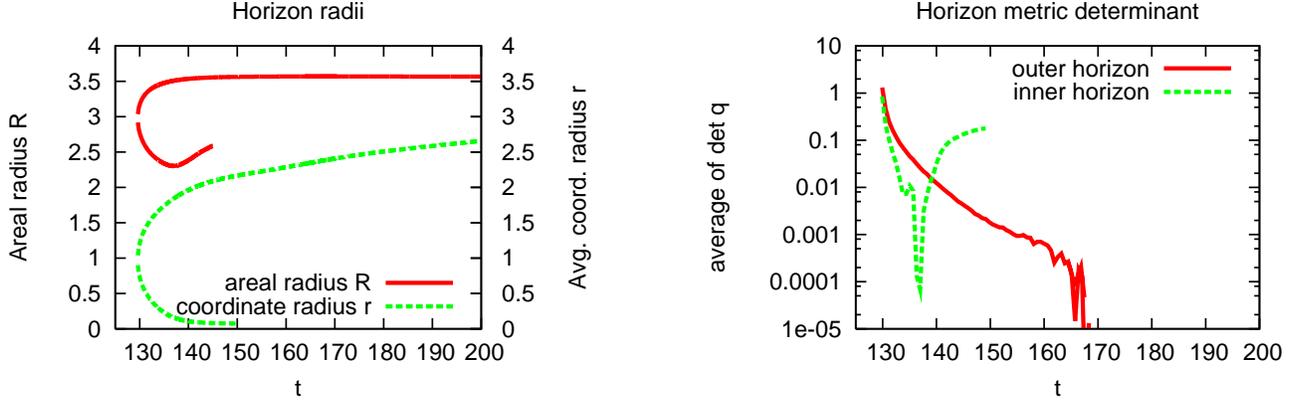}
  \caption{The average coordinate radius and the area radius as a
    function of time for the outer and inner MTTs for the neutron star
    collapse.  The inner horizon is not to be trusted after $t \approx
    140$ due to lack of resolution, since its coordinate radius has
    become very small by that time.}
  \label{fig:rns-radius}
\end{figure}

We find an apparent horizon starting at about $t=130$; this time is
mainly dependent on the details of how the collapse is induced and has
no intrinsic meaning.  The horizon is born with an irreducible mass of
about $M_\mathrm{irr}=1.51$ and an angular momentum of $J=0.89$
($a=0.38$), giving it a total mass of $M_H=1.54$.  Some time after
$t=185$, a singularity forms in the spacetime, and the simulation
aborts because we do not use excision inside the apparent horizon.  As
before, a pair of MOTSs is formed, an outer and an inner one.  The
outer MTT is spacelike, has increasing area, and tends to null at late
times.  In this case, the inner MTT remains spacelike.  However, its
area decreases because we are traversing it in the inward direction;
in other words, the time evolution vector $t^a$ is such that at the
inner MTT, $t\cdot \hat{r} < 0$ so that that the area decreases along
$t^a$.  Our gauge conditions are such that the outer horizon grows in
coordinate space while the inner horizon shrinks.  After about
$t=140$, the inner horizon is so small that we do not have enough
resolution to track it beyond that time.  See figure
\ref{fig:rns-radius}.  The areal radius of the outer MTT increases but
not as rapidly as the coordinate radius; it levels off at later times.
The area radius of the inner horizon decreases initially and shows an
increase at later times, but this is probably just a numerical artefact
due to poor resolution at later times.

\begin{figure} 
  \includegraphics[width=0.45\textwidth]{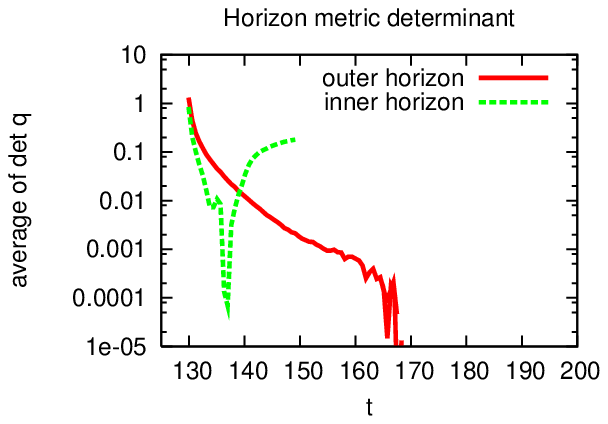}
  \includegraphics[width=0.45\textwidth]{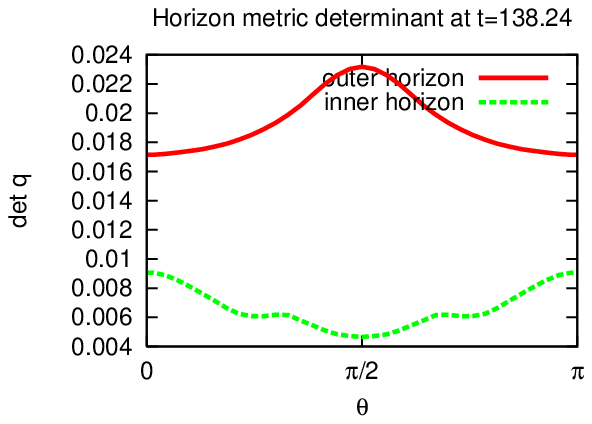}
  \caption{Average of the determinant of the horizon world tube's
    three metric vs.\ time, and vs.\ latitude $\theta$ at $t=138.24$
    for the inner and outer horizons for the neutron star collapse. }
  \label{fig:rns-3det} 
\end{figure}

Figure \ref{fig:rns-3det} shows the determinant of the metric on the
MTTs.  The outer MTT is initially spacelike, which is consistent with
its growing, and exponentially approaches null at late times. After
about $t=160$, the simulation cannot distinguish the horizon world
tube signature from null any more.  As an example we also show the
determinant as a function of the latitude $\theta$ at $t \approx 138$,
and the
average value of the determinant over the horizon as a function of
time.  The inner MTT is also spacelike and becomes more and more
null at least as long as we are able to track it reliably.

\begin{figure} 
  \includegraphics[width=0.45\textwidth]{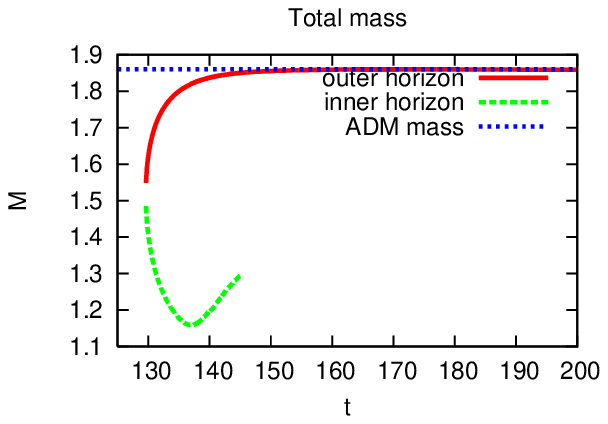}
  \includegraphics[width=0.45\textwidth]{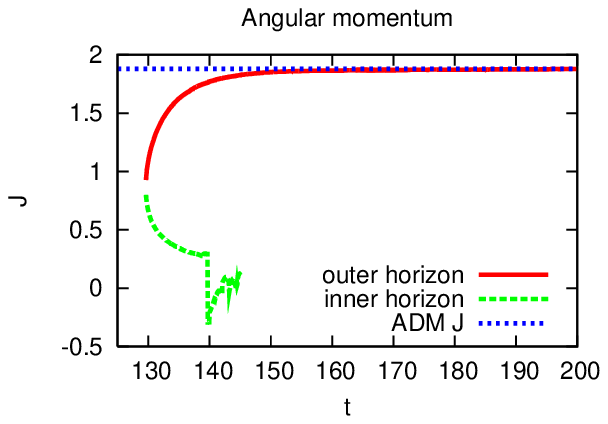}
  \caption{The total mass $M_H$,, and angular momentum $J$ as a
    function of time for the outer and inner MTTs for the neutron star
    collapse.}
  \label{fig:rns-masses}
\end{figure}

Figure \ref{fig:rns-masses} shows the outer horizon has grown at
$t=155$ to an irreducible mass of $M_\mathrm{irr}=1.80$ and an angular
momentum of $J=1.93$ ($a=0.55$), giving it a total mass of $M_H=1.87$.
For comparison, the corresponding ADM quantities are
$M_\mathrm{ADM}=1.86$ and $J_\ADM=1.88$ ($a=0.54$).  Because the
spacetime is axially symmetric, gravitational waves cannot carry away
angular momentum.  That means that the spin $a=J/M^2$ is approximately
correct at late times.
Unlike in the
non-axisymmetric black hole collision discussed earlier, the present
case is explicitly axisymmetric and there are no problems with
locating the rotational symmetry vector.

\begin{figure} 
  \includegraphics[width=0.45\textwidth]{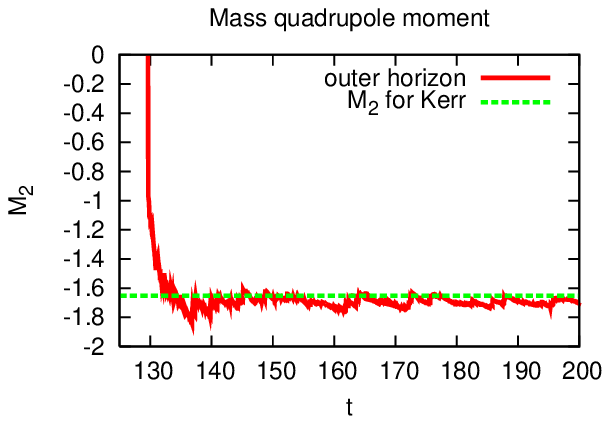}
  \includegraphics[width=0.45\textwidth]{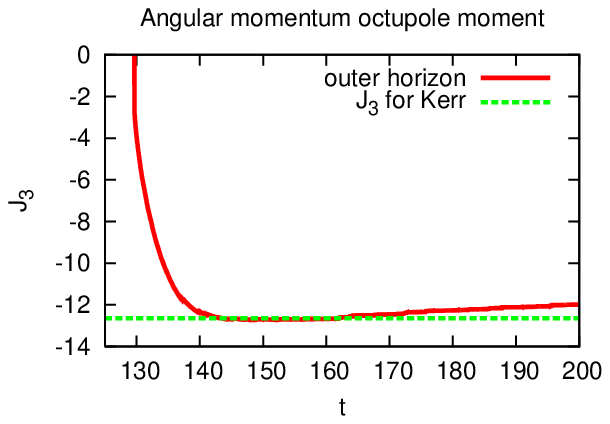}
  \caption{Horizon mass quadrupole moment $M_2$ and angular momentum
    octopole moment $J_3$ vs.\ time for the neutron star collapse.
    For comparison, the values for a Kerr black hole with the same ADM
    mass and angular momentum as the initial data are also shown. }
  \label{fig:rns-multipoles}
\end{figure}

Figure \ref{fig:rns-multipoles} shows the mass quadrupole moment $M_2$
and the angular momentum octopole moment of the outer and inner MTTs
as a function of time.  Given that we know the asymptotic values of
the area and angular momentum of these MTTs (the ADM values), we can
also calculate the \emph{expected} values of $M_2$ and $J_3$ at late
times.  The plots clearly show that the values of $M_2$ and $J_3$
approach the Kerr values at later times (though this matching is not
exact, presumably due to numerical errors).
Also note that $M_2$ is noisy.  We have observed such noise only in
simulations that include matter, and we find that this noise is much
improved by using artificial dissipation on the spacetime variables
(which we do).  The angular momentum multipoles seem unaffected.

\section{Discussion}
\label{sec:conclusion}

In this article, we have applied the dynamical horizon formalism to
numerical simulations of black hole spacetimes. The main theme in this
formalism is to take trapped surfaces seriously as a way of describing
black hole physics.  Marginally trapped surfaces behave more regularly
that one might have expected previously, and they are useful for
extracting interesting physical information about the horizon. We have
shown how the mass, angular momentum, multipole moments, and the flux
of energy due to in-falling gravitational radiation and matter can be
calculated in a coordinate independent way (given a particular time
slicing of our spacetime). We have implemented these ideas numerically
and shown three concrete examples.  In these examples, we see how the
black hole is formed, how it grows, and how it settles down to an
isolated Kerr black hole.  We have also seen that the dynamical
horizon formalism is valuable for exploring the geometry of the
trapped region.  It allows us to classify various types of trapped
surfaces which might appear during the course of a gravitational
collapse or a black hole coalescence.  Finally, these ideas can also
be viewed as a set of diagnostic tools which allow us to keep track of
what is going on during the course of a numerical simulation, and
whether numerical results make sense and satisfy some basic, but
non-trivial properties in the strong field region.

Some suggestions for future work:

\begin{description}

\item[i.] As mentioned in the text, the calculation of the axial
  vector $\varphi^a$ for non-axisymmetric cases is not yet
  satisfactory.  We have used the method suggested in \cite{Dreyer02a}
  which works well enough at early and late times, when the horizon is
  approximately axisymmetric.  However, in general, the result is not
  guaranteed to be divergence free and thus the angular momentum not
  guaranteed to be gauge invariant. We have not yet implemented the
  generalization described in section \ref{subsec:angmom}
  satisfactorily; this is work in progress.

\item[ii.]  The accuracy of the numerical examples that we have shown
  decreases with time, and this is a common feature of most present
  day black hole numerical simulations.  Thus, we have not been
  conclusively able to prove that the black hole settles down to Kerr
  (though there are strong indications that this does happen). We have
  not been able to extract the rate at which equilibrium is reached,
  thereby extending Price's law (see \cite{Price72} and e.g.\
  \cite{Dafermos05}) to more general situations, but this is, in
  principle possible and requires more stable and accurate
  simulations.  Similarly, we have not been able to accurately
  calculate the asymptotic value of the black hole mass $M^\infty$.
  The difference $M_\ADM -M^\infty$ is, in principle, a reliable
  estimate of the amount of energy radiated to infinity.  While the
  ADM mass is hard to calculate reliably during the simulation because
  of the finite grid and low resolution in the asymptotic region, it
  can usually be calculated accurately from the initial data themself.
  Calculating $M^\infty$ and understanding this estimate of the
  radiated energy requires more accurate and stable runs, applied to
  diverse and realistic initial data.  The results of \cite{Booth04a}
  could also be used to study the approach to equilibrium.  

\item[iii.] It would be useful numerically to have a gauge condition
  which ensures that the horizon stays at the same coordinate location
  at all times.  While such conditions are not difficult to find in
  the isolated case, dynamical situations are harder.  Given the
  location of an outer MOTS at a particular instant of time, the
  results and methods of \cite{ams05} can be used to \emph{predict}
  the location of the MOTS at the next instant by solving an elliptic
  equation on the MOTS.  This could be used to construct appropriate
  gauge conditions and evolution schemes which take the horizon
  geometry into account \cite{Anninos94e, Eardley98}.

\item[iv.] What happens to the inner horizon of figures
  \ref{fig:multibh}, \ref{fig:headon-shape-t1}, and
  \ref{fig:nonheadon-shape-t18}?  As described in section
  \ref{subsec:resultsbl}, the eventual fate of these inner MTTs and
  the two individual horizons is not yet known, and would be
  interesting to investigate further.  This requires simulations with
  higher resolution near the inner horizons, different gauge
  conditions, and perhaps also AH trackers capable of handling
  non-star-shaped surfaces, and perhaps also higher genus surfaces.

\item[v.] Can the methods of \cite{ams05} be extended for MOTSs which
  are not strictly-stably-outermost?  In this regard, it would be
  interesting to study the stability operator $L_\Sigma$ introduced in
  \cite{ams05}. For a strictly-stably-outermost MOTS, the principle
  eigenvalue of $L_\Sigma$ turns out to be strictly positive and this
  is an important ingredient in the existence results.  A numerical
  computation of the eigenvalues of this operator, especially during
  the transition between inner and outer MTTs and for the inner
  non-spacelike MTTs might lead to further insights.

\end{description}

\vspace{2ex}

\begin{acknowledgments}
  We are grateful to Lars Andersson and Abhay Ashtekar for many
  valuable suggestions and fruitful discussions.  We also thank Ivan
  Booth, Sergio Dain, Steve Fairhurst, Greg Galloway, Ian Hawke, Sean
  Hayward, Jan Metzger, Denis Pollney, Reinhard Prix, Jonathan
  Thornburg, and Robert Wald for useful discussions.

  As always, our numerical calculations would have been impossible
  without the large number of people who made their work available to
  the public: we used the Cactus framework \cite{Goodale02a,
    cactusweb1} and the \texttt{CactusEinstein} infrastructure
  \cite{cactuseinsteinweb} with a number of locally developed thorns,
  such as the initial data solver \texttt{TwoPunctures} by M. Ansorg,
  the mesh refinement criteria set up via \texttt{WhiskyCarpetRegrid}
  by C. D. Ott and I. Hawke, and the horizon finder
  \texttt{AHFinderDirect} by J. Thornburg.  We also used the general
  relativistic hydrodynamics code \texttt{Whisky} \cite{whiskyweb}
  developed by the authors of \cite{Baiotti04}, and the initial data
  generator \texttt{RNSID} by N. Stergioulas, which were both
  developed during the EU training network ``Sources of Gravitational
  Waves''.  The code uses routines of the \texttt{LAPACK} \cite{laug,
    lapackweb} and \texttt{BLAS} \cite{blasweb} libraries from the
  Netlib Repository \cite{netlibweb}, the Numerical Recipes
  \cite{Press86}, and the \texttt{UMFPACK} \cite{umfpackweb} library.
  The numerical simulations were performed on the \emph{Peyote}
  Beowulf Cluster at the AEI.  ES was partly funded by the DFG's special
  research centre SFB TR/7 ``Gravitational Wave Astronomy''.  This
  work was supported by the Albert--Einstein--Institut and the Center
  for Computation \& Technology at LSU.
\end{acknowledgments}

\bibliographystyle{bibtex/apsrev-titles}


\bibliography{bibtex/references}

\begin{thebibliography}{89}
\expandafter\ifx\csname natexlab\endcsname\relax\def\natexlab#1{#1}\fi
\expandafter\ifx\csname bibnamefont\endcsname\relax
  \def\bibnamefont#1{#1}\fi
\expandafter\ifx\csname bibfnamefont\endcsname\relax
  \def\bibfnamefont#1{#1}\fi
\expandafter\ifx\csname citenamefont\endcsname\relax
  \def\citenamefont#1{#1}\fi
\expandafter\ifx\csname url\endcsname\relax
  \def\url#1{\texttt{#1}}\fi
\expandafter\ifx\csname urlprefix\endcsname\relax\def\urlprefix{URL }\fi
\providecommand{\bibinfo}[2]{#2}
\providecommand{\eprint}[2][]{\url{#2}}

\bibitem[{\citenamefont{Pretorius}(2005)}]{Pretorius:2005gq}
\bibinfo{author}{\bibfnamefont{F.}~\bibnamefont{Pretorius}},
  \emph{\bibinfo{title}{Evolution of binary black hole spacetimes}},
  \bibinfo{journal}{Phys. Rev. Lett.} \textbf{\bibinfo{volume}{95}},
  \bibinfo{pages}{121101} (\bibinfo{year}{2005}), \eprint{gr-qc/0507014}.

\bibitem[{\citenamefont{Campanelli
  et~al.}(2006{\natexlab{a}})\citenamefont{Campanelli, Lousto, Marronetti, and
  Zlochower}}]{Campanelli:2005dd}
\bibinfo{author}{\bibfnamefont{M.}~\bibnamefont{Campanelli}},
  \bibinfo{author}{\bibfnamefont{C.~O.} \bibnamefont{Lousto}},
  \bibinfo{author}{\bibfnamefont{P.}~\bibnamefont{Marronetti}},
  \bibnamefont{and}
  \bibinfo{author}{\bibfnamefont{Y.}~\bibnamefont{Zlochower}},
  \emph{\bibinfo{title}{Accurate evolutions of orbiting black-hole binaries
  without excision}}, \bibinfo{journal}{Phys. Rev. Letter}
  \textbf{\bibinfo{volume}{96}}, \bibinfo{pages}{111101}
  (\bibinfo{year}{2006}{\natexlab{a}}), \eprint{gr-qc/0511048}.

\bibitem[{\citenamefont{Baker et~al.}(2006{\natexlab{a}})\citenamefont{Baker,
  Centrella, Choi, Koppitz, and van Meter}}]{Baker05a}
\bibinfo{author}{\bibfnamefont{J.~G.} \bibnamefont{Baker}},
  \bibinfo{author}{\bibfnamefont{J.}~\bibnamefont{Centrella}},
  \bibinfo{author}{\bibfnamefont{D.-I.} \bibnamefont{Choi}},
  \bibinfo{author}{\bibfnamefont{M.}~\bibnamefont{Koppitz}}, \bibnamefont{and}
  \bibinfo{author}{\bibfnamefont{J.}~\bibnamefont{van Meter}},
  \emph{\bibinfo{title}{Gravitational wave extraction from an inspiraling
  configuration of merging black holes}}, \bibinfo{journal}{Phys. Rev. Lett.}
  \textbf{\bibinfo{volume}{96}}, \bibinfo{pages}{111102}
  (\bibinfo{year}{2006}{\natexlab{a}}), \eprint{gr-qc/0511103}.

\bibitem[{\citenamefont{Herrmann et~al.}(2006)\citenamefont{Herrmann,
  Shoemaker, and Laguna}}]{Herrmann:2006ks}
\bibinfo{author}{\bibfnamefont{F.}~\bibnamefont{Herrmann}},
  \bibinfo{author}{\bibfnamefont{D.}~\bibnamefont{Shoemaker}},
  \bibnamefont{and} \bibinfo{author}{\bibfnamefont{P.}~\bibnamefont{Laguna}},
  \emph{\bibinfo{title}{Unequal-mass binary black hole inspirals}}
  (\bibinfo{year}{2006}), \eprint{gr-qc/0601026}.

\bibitem[{\citenamefont{Campanelli
  et~al.}(2006{\natexlab{b}})\citenamefont{Campanelli, Lousto, and
  Zlochower}}]{Campanelli:2006gf}
\bibinfo{author}{\bibfnamefont{M.}~\bibnamefont{Campanelli}},
  \bibinfo{author}{\bibfnamefont{C.~O.} \bibnamefont{Lousto}},
  \bibnamefont{and}
  \bibinfo{author}{\bibfnamefont{Y.}~\bibnamefont{Zlochower}},
  \emph{\bibinfo{title}{The last orbit of binary black holes}},
  \bibinfo{journal}{Phys. Rev. D} \textbf{\bibinfo{volume}{73}},
  \bibinfo{pages}{061501(R)} (\bibinfo{year}{2006}{\natexlab{b}}),
  \eprint{gr-qc/0601091}.

\bibitem[{\citenamefont{Baker et~al.}(2006{\natexlab{b}})\citenamefont{Baker,
  Centrella, Choi, Koppitz, and van Meter}}]{Baker:2006yw}
\bibinfo{author}{\bibfnamefont{J.~G.} \bibnamefont{Baker}},
  \bibinfo{author}{\bibfnamefont{J.}~\bibnamefont{Centrella}},
  \bibinfo{author}{\bibfnamefont{D.-I.} \bibnamefont{Choi}},
  \bibinfo{author}{\bibfnamefont{M.}~\bibnamefont{Koppitz}}, \bibnamefont{and}
  \bibinfo{author}{\bibfnamefont{J.}~\bibnamefont{van Meter}},
  \emph{\bibinfo{title}{Binary black hole merger dynamics and waveforms}},
  \bibinfo{journal}{Phys. Rev. D} \textbf{\bibinfo{volume}{73}},
  \bibinfo{pages}{104002} (\bibinfo{year}{2006}{\natexlab{b}}).

\bibitem[{\citenamefont{Diener et~al.}(2006)\citenamefont{Diener, Herrmann,
  Pollney, Schnetter, Seidel, Takahashi, Thornburg, and
  Ventrella}}]{Diener-etal-2006a}
\bibinfo{author}{\bibfnamefont{P.}~\bibnamefont{Diener}},
  \bibinfo{author}{\bibfnamefont{F.}~\bibnamefont{Herrmann}},
  \bibinfo{author}{\bibfnamefont{D.}~\bibnamefont{Pollney}},
  \bibinfo{author}{\bibfnamefont{E.}~\bibnamefont{Schnetter}},
  \bibinfo{author}{\bibfnamefont{E.}~\bibnamefont{Seidel}},
  \bibinfo{author}{\bibfnamefont{R.}~\bibnamefont{Takahashi}},
  \bibinfo{author}{\bibfnamefont{J.}~\bibnamefont{Thornburg}},
  \bibnamefont{and}
  \bibinfo{author}{\bibfnamefont{J.}~\bibnamefont{Ventrella}},
  \emph{\bibinfo{title}{Accurate evolution of orbiting binary black holes}},
  \bibinfo{journal}{Phys. Rev. Lett.} \textbf{\bibinfo{volume}{96}},
  \bibinfo{pages}{121101} (\bibinfo{year}{2006}), \eprint{gr-qc/0512108},
  \urlprefix\url{http://link.aps.org/abstract/PRL/v96/e121101}.

\bibitem[{\citenamefont{Campanelli
  et~al.}(2006{\natexlab{c}})\citenamefont{Campanelli, Lousto, and
  Zlochower}}]{Campanelli:2006uy}
\bibinfo{author}{\bibfnamefont{M.}~\bibnamefont{Campanelli}},
  \bibinfo{author}{\bibfnamefont{C.~O.} \bibnamefont{Lousto}},
  \bibnamefont{and}
  \bibinfo{author}{\bibfnamefont{Y.}~\bibnamefont{Zlochower}},
  \emph{\bibinfo{title}{Gravitational radiation from spinning-black-hole
  binaries: The orbital hang up}} (\bibinfo{year}{2006}{\natexlab{c}}),
  \eprint{gr-qc/0604012}.

\bibitem[{\citenamefont{Ashtekar and Krishnan}(2002)}]{Ashtekar02a}
\bibinfo{author}{\bibfnamefont{A.}~\bibnamefont{Ashtekar}} \bibnamefont{and}
  \bibinfo{author}{\bibfnamefont{B.}~\bibnamefont{Krishnan}},
  \emph{\bibinfo{title}{{Dynamical} {Horizons}: Energy, angular momentum,
  fluxes, and balance laws}}, \bibinfo{journal}{Phys. Rev. Lett.}
  \textbf{\bibinfo{volume}{89}}, \bibinfo{pages}{261101}
  (\bibinfo{year}{2002}), \eprint{gr-qc/0207080}.

\bibitem[{\citenamefont{Ashtekar and Krishnan}(2003)}]{Ashtekar03a}
\bibinfo{author}{\bibfnamefont{A.}~\bibnamefont{Ashtekar}} \bibnamefont{and}
  \bibinfo{author}{\bibfnamefont{B.}~\bibnamefont{Krishnan}},
  \emph{\bibinfo{title}{Dynamical horizons and their properties}},
  \bibinfo{journal}{Phys. Rev. D} \textbf{\bibinfo{volume}{68}},
  \bibinfo{pages}{104030} (\bibinfo{year}{2003}), \eprint{gr-qc/0308033}.

\bibitem[{\citenamefont{Ashtekar et~al.}(1999)\citenamefont{Ashtekar, Beetle,
  and Fairhurst}}]{Ashtekar98a}
\bibinfo{author}{\bibfnamefont{A.}~\bibnamefont{Ashtekar}},
  \bibinfo{author}{\bibfnamefont{C.}~\bibnamefont{Beetle}}, \bibnamefont{and}
  \bibinfo{author}{\bibfnamefont{S.}~\bibnamefont{Fairhurst}},
  \emph{\bibinfo{title}{Isolated horizons: A generalization of black hole
  mechanics}}, \bibinfo{journal}{Class. Quantum Grav.}
  \textbf{\bibinfo{volume}{16}}, \bibinfo{pages}{L1} (\bibinfo{year}{1999}),
  \eprint{gr-qc/9812065}.

\bibitem[{\citenamefont{Ashtekar
  et~al.}(2000{\natexlab{a}})\citenamefont{Ashtekar, Beetle, and
  Fairhurst}}]{Ashtekar99a}
\bibinfo{author}{\bibfnamefont{A.}~\bibnamefont{Ashtekar}},
  \bibinfo{author}{\bibfnamefont{C.}~\bibnamefont{Beetle}}, \bibnamefont{and}
  \bibinfo{author}{\bibfnamefont{S.}~\bibnamefont{Fairhurst}},
  \emph{\bibinfo{title}{Mechanics of isolated horizons}},
  \bibinfo{journal}{Class. Quantum Grav.} \textbf{\bibinfo{volume}{17}},
  \bibinfo{pages}{253} (\bibinfo{year}{2000}{\natexlab{a}}),
  \eprint{gr-qc/9907068}.

\bibitem[{\citenamefont{Ashtekar et~al.}(2001)\citenamefont{Ashtekar, Beetle,
  and Lewandowski}}]{Ashtekar01a}
\bibinfo{author}{\bibfnamefont{A.}~\bibnamefont{Ashtekar}},
  \bibinfo{author}{\bibfnamefont{C.}~\bibnamefont{Beetle}}, \bibnamefont{and}
  \bibinfo{author}{\bibfnamefont{J.}~\bibnamefont{Lewandowski}},
  \emph{\bibinfo{title}{Mechanics of rotating isolated horizons}},
  \bibinfo{journal}{Phys. Rev. D} \textbf{\bibinfo{volume}{64}},
  \bibinfo{pages}{044016} (\bibinfo{year}{2001}), \eprint{gr-qc/0103026}.

\bibitem[{\citenamefont{Ashtekar
  et~al.}(2000{\natexlab{b}})\citenamefont{Ashtekar, Fairhurst, and
  Krishnan}}]{Ashtekar00b}
\bibinfo{author}{\bibfnamefont{A.}~\bibnamefont{Ashtekar}},
  \bibinfo{author}{\bibfnamefont{S.}~\bibnamefont{Fairhurst}},
  \bibnamefont{and} \bibinfo{author}{\bibfnamefont{B.}~\bibnamefont{Krishnan}},
  \emph{\bibinfo{title}{Isolated horizons: {H}amiltonian evolution and the
  first law}}, \bibinfo{journal}{Phys. Rev. D} \textbf{\bibinfo{volume}{62}},
  \bibinfo{pages}{104025} (\bibinfo{year}{2000}{\natexlab{b}}),
  \eprint{gr-qc/0005083}.

\bibitem[{\citenamefont{Ashtekar
  et~al.}(2000{\natexlab{c}})\citenamefont{Ashtekar, Beetle, Dreyer, Fairhurst,
  Krishnan, Lewandowski, and Wisniewski}}]{Ashtekar00a}
\bibinfo{author}{\bibfnamefont{A.}~\bibnamefont{Ashtekar}},
  \bibinfo{author}{\bibfnamefont{C.}~\bibnamefont{Beetle}},
  \bibinfo{author}{\bibfnamefont{O.}~\bibnamefont{Dreyer}},
  \bibinfo{author}{\bibfnamefont{S.}~\bibnamefont{Fairhurst}},
  \bibinfo{author}{\bibfnamefont{B.}~\bibnamefont{Krishnan}},
  \bibinfo{author}{\bibfnamefont{J.}~\bibnamefont{Lewandowski}},
  \bibnamefont{and}
  \bibinfo{author}{\bibfnamefont{J.}~\bibnamefont{Wisniewski}},
  \emph{\bibinfo{title}{Generic isolated horizons and their applications}},
  \bibinfo{journal}{Phys. Rev. Lett.} \textbf{\bibinfo{volume}{85}},
  \bibinfo{pages}{3564} (\bibinfo{year}{2000}{\natexlab{c}}),
  \eprint{gr-qc/0006006}.

\bibitem[{\citenamefont{Hayward}(1994{\natexlab{a}})}]{Hayward94a}
\bibinfo{author}{\bibfnamefont{S.~A.} \bibnamefont{Hayward}},
  \emph{\bibinfo{title}{General laws of black hole dynamics}},
  \bibinfo{journal}{Phys. Rev. D} \textbf{\bibinfo{volume}{49}},
  \bibinfo{pages}{6467} (\bibinfo{year}{1994}{\natexlab{a}}),
  \eprint{gr-qc/9306006},
  \urlprefix\url{http://link.aps.org/abstract/PRD/v49/p6467}.

\bibitem[{\citenamefont{Hayward}(1994{\natexlab{b}})}]{Hayward94b}
\bibinfo{author}{\bibfnamefont{S.}~\bibnamefont{Hayward}},
  \emph{\bibinfo{title}{Spin-coefficient form of the new laws of black hole
  dynamics}}, \bibinfo{journal}{Class. Quantum Grav.}
  \textbf{\bibinfo{volume}{11}}, \bibinfo{pages}{3025}
  (\bibinfo{year}{1994}{\natexlab{b}}), \eprint{gr-qc/9406033}.

\bibitem[{\citenamefont{Hayward}(2004)}]{Hayward04}
\bibinfo{author}{\bibfnamefont{S.~A.} \bibnamefont{Hayward}},
  \emph{\bibinfo{title}{Energy and entropy conservation for dynamical black
  holes}}, \bibinfo{journal}{Phys. Rev. D} \textbf{\bibinfo{volume}{70}},
  \bibinfo{pages}{104027} (\bibinfo{year}{2004}), \eprint{gr-qc/0408008}.

\bibitem[{\citenamefont{Ashtekar and Krishnan}(2004)}]{Ashtekar:2004cn}
\bibinfo{author}{\bibfnamefont{A.}~\bibnamefont{Ashtekar}} \bibnamefont{and}
  \bibinfo{author}{\bibfnamefont{B.}~\bibnamefont{Krishnan}},
  \emph{\bibinfo{title}{Isolated and dynamical horizons and their
  applications}}, \bibinfo{journal}{Living Rev. Rel.}
  \textbf{\bibinfo{volume}{7}}, \bibinfo{pages}{10} (\bibinfo{year}{2004}),
  \eprint{gr-qc/0407042}.

\bibitem[{\citenamefont{Booth}(2005)}]{Booth-review}
\bibinfo{author}{\bibfnamefont{I.}~\bibnamefont{Booth}},
  \emph{\bibinfo{title}{Black hole boundaries}}, \bibinfo{journal}{Can. J.
  Phys.} \textbf{\bibinfo{volume}{83}}, \bibinfo{pages}{1073}
  (\bibinfo{year}{2005}), \eprint{gr-qc/0508107}.

\bibitem[{\citenamefont{Gourgoulhon and
  Jaramillo}(2006)}]{Gourgoulhon-Jaramillo-Review}
\bibinfo{author}{\bibfnamefont{E.}~\bibnamefont{Gourgoulhon}} \bibnamefont{and}
  \bibinfo{author}{\bibfnamefont{J.~L.} \bibnamefont{Jaramillo}},
  \emph{\bibinfo{title}{A $3+1$ perspective on null hypersurfaces and isolated
  horizons}}, \bibinfo{journal}{Physics Reports}
  \textbf{\bibinfo{volume}{423}}, \bibinfo{pages}{159} (\bibinfo{year}{2006}),
  \eprint{gr-qc/0503113}.

\bibitem[{\citenamefont{Diener}(2003)}]{Diener03a}
\bibinfo{author}{\bibfnamefont{P.}~\bibnamefont{Diener}},
  \emph{\bibinfo{title}{A new general purpose event horizon finder for {3D}
  numerical spacetimes}}, \bibinfo{journal}{Class. Quantum Grav.}
  \textbf{\bibinfo{volume}{20}}, \bibinfo{pages}{4901} (\bibinfo{year}{2003}),
  \eprint{gr-qc/0305039},
  \urlprefix\url{http://stacks.iop.org/0264-9381/20/4901}.

\bibitem[{\citenamefont{Hughes et~al.}(1994)\citenamefont{Hughes, Keeton,
  Walker, Walsh, Shapiro, and Teukolsky}}]{Hughes94a}
\bibinfo{author}{\bibfnamefont{S.~A.} \bibnamefont{Hughes}},
  \bibinfo{author}{\bibfnamefont{C.~R.} \bibnamefont{Keeton},
  \bibfnamefont{II}}, \bibinfo{author}{\bibfnamefont{P.}~\bibnamefont{Walker}},
  \bibinfo{author}{\bibfnamefont{K.~T.} \bibnamefont{Walsh}},
  \bibinfo{author}{\bibfnamefont{S.~L.} \bibnamefont{Shapiro}},
  \bibnamefont{and} \bibinfo{author}{\bibfnamefont{S.~A.}
  \bibnamefont{Teukolsky}}, \emph{\bibinfo{title}{Finding black holes in
  numerical spacetimes}}, \bibinfo{journal}{Phys. Rev. D}
  \textbf{\bibinfo{volume}{49}}, \bibinfo{pages}{4004} (\bibinfo{year}{1994}),
  \urlprefix\url{http://link.aps.org/abstract/PRD/v49/p4004}.

\bibitem[{\citenamefont{Anninos et~al.}(1994)\citenamefont{Anninos, Bernstein,
  Brandt, Hobill, Seidel, and Smarr}}]{Anninos93a}
\bibinfo{author}{\bibfnamefont{P.}~\bibnamefont{Anninos}},
  \bibinfo{author}{\bibfnamefont{D.}~\bibnamefont{Bernstein}},
  \bibinfo{author}{\bibfnamefont{S.~R.} \bibnamefont{Brandt}},
  \bibinfo{author}{\bibfnamefont{D.}~\bibnamefont{Hobill}},
  \bibinfo{author}{\bibfnamefont{E.}~\bibnamefont{Seidel}}, \bibnamefont{and}
  \bibinfo{author}{\bibfnamefont{L.}~\bibnamefont{Smarr}},
  \emph{\bibinfo{title}{Dynamics of black hole apparent horizons}},
  \bibinfo{journal}{Phys. Rev. D} \textbf{\bibinfo{volume}{50}},
  \bibinfo{pages}{3801} (\bibinfo{year}{1994}),
  \urlprefix\url{http://link.aps.org/abstract/PRD/v50/p3801}.

\bibitem[{\citenamefont{Anninos
  et~al.}(1995{\natexlab{a}})\citenamefont{Anninos, Bernstein, Brandt, Libson,
  Mass{\'o}, Seidel, Smarr, Suen, and Walker}}]{Anninos94f}
\bibinfo{author}{\bibfnamefont{P.}~\bibnamefont{Anninos}},
  \bibinfo{author}{\bibfnamefont{D.}~\bibnamefont{Bernstein}},
  \bibinfo{author}{\bibfnamefont{S.}~\bibnamefont{Brandt}},
  \bibinfo{author}{\bibfnamefont{J.}~\bibnamefont{Libson}},
  \bibinfo{author}{\bibfnamefont{J.}~\bibnamefont{Mass{\'o}}},
  \bibinfo{author}{\bibfnamefont{E.}~\bibnamefont{Seidel}},
  \bibinfo{author}{\bibfnamefont{L.}~\bibnamefont{Smarr}},
  \bibinfo{author}{\bibfnamefont{W.-M.} \bibnamefont{Suen}}, \bibnamefont{and}
  \bibinfo{author}{\bibfnamefont{P.}~\bibnamefont{Walker}},
  \emph{\bibinfo{title}{Dynamics of apparent and event horizons}},
  \bibinfo{journal}{Phys. Rev. Lett.} \textbf{\bibinfo{volume}{74}},
  \bibinfo{pages}{630} (\bibinfo{year}{1995}{\natexlab{a}}),
  \eprint{gr-qc/9403011},
  \urlprefix\url{http://link.aps.org/abstract/PRL/v74/p630}.

\bibitem[{\citenamefont{Brandt and Seidel}(1995)}]{Brandt94c}
\bibinfo{author}{\bibfnamefont{S.~R.} \bibnamefont{Brandt}} \bibnamefont{and}
  \bibinfo{author}{\bibfnamefont{E.}~\bibnamefont{Seidel}},
  \emph{\bibinfo{title}{Evolution of distorted rotating black holes {II}:
  Dynamics and analysis}}, \bibinfo{journal}{Phys. Rev. D}
  \textbf{\bibinfo{volume}{52}}, \bibinfo{pages}{870} (\bibinfo{year}{1995}),
  \urlprefix\url{http://link.aps.org/abstract/PRD/v52/p870}.

\bibitem[{\citenamefont{Libson et~al.}(1996)\citenamefont{Libson, Mass{\'o},
  Seidel, Suen, and Walker}}]{Libson94a}
\bibinfo{author}{\bibfnamefont{J.}~\bibnamefont{Libson}},
  \bibinfo{author}{\bibfnamefont{J.}~\bibnamefont{Mass{\'o}}},
  \bibinfo{author}{\bibfnamefont{E.}~\bibnamefont{Seidel}},
  \bibinfo{author}{\bibfnamefont{W.-M.} \bibnamefont{Suen}}, \bibnamefont{and}
  \bibinfo{author}{\bibfnamefont{P.}~\bibnamefont{Walker}},
  \emph{\bibinfo{title}{Event horizons in numerical relativity: Methods and
  tests}}, \bibinfo{journal}{Phys. Rev. D} \textbf{\bibinfo{volume}{53}},
  \bibinfo{pages}{4335} (\bibinfo{year}{1996}),
  \urlprefix\url{http://link.aps.org/abstract/PRD/v53/p4335}.

\bibitem[{\citenamefont{Mass{\'o} et~al.}(1999)\citenamefont{Mass{\'o}, Seidel,
  Suen, and Walker}}]{Masso95a}
\bibinfo{author}{\bibfnamefont{J.}~\bibnamefont{Mass{\'o}}},
  \bibinfo{author}{\bibfnamefont{E.}~\bibnamefont{Seidel}},
  \bibinfo{author}{\bibfnamefont{W.-M.} \bibnamefont{Suen}}, \bibnamefont{and}
  \bibinfo{author}{\bibfnamefont{P.}~\bibnamefont{Walker}},
  \emph{\bibinfo{title}{Event horizons in numerical relativity {II}: Analyzing
  the horizon}}, \bibinfo{journal}{Phys. Rev. D} \textbf{\bibinfo{volume}{59}},
  \bibinfo{pages}{064015} (\bibinfo{year}{1999}),
  \bibinfo{note}{gr-qc/9804059},
  \urlprefix\url{http://link.aps.org/abstract/PRD/v59/e064015}.

\bibitem[{\citenamefont{Wald}(1984)}]{Wald84}
\bibinfo{author}{\bibfnamefont{R.~M.} \bibnamefont{Wald}},
  \emph{\bibinfo{title}{General relativity}} (\bibinfo{publisher}{The
  University of Chicago Press}, \bibinfo{address}{Chicago},
  \bibinfo{year}{1984}), ISBN \bibinfo{isbn}{0-226-87032-4 (hardcover),
  0-226-87033-2 (paperback)}.

\bibitem[{\citenamefont{Penrose}(1965)}]{Penrose65}
\bibinfo{author}{\bibfnamefont{R.}~\bibnamefont{Penrose}},
  \emph{\bibinfo{title}{Gravitational collapse and space-time singularities}},
  \bibinfo{journal}{Phys. Rev. Lett.} \textbf{\bibinfo{volume}{14}},
  \bibinfo{pages}{57} (\bibinfo{year}{1965}).

\bibitem[{\citenamefont{Penrose and Hawking}(1970)}]{Penrose70a}
\bibinfo{author}{\bibfnamefont{R.}~\bibnamefont{Penrose}} \bibnamefont{and}
  \bibinfo{author}{\bibfnamefont{S.~W.} \bibnamefont{Hawking}},
  \emph{\bibinfo{title}{The singularities of gravitational collapse and
  cosmology}}, \bibinfo{journal}{Proc. Roy. Soc. Lond. A}
  \textbf{\bibinfo{volume}{314}}, \bibinfo{pages}{529} (\bibinfo{year}{1970}).

\bibitem[{\citenamefont{Hawking and Ellis}(1973)}]{Hawking73a}
\bibinfo{author}{\bibfnamefont{S.~W.} \bibnamefont{Hawking}} \bibnamefont{and}
  \bibinfo{author}{\bibfnamefont{G.~F.~R.} \bibnamefont{Ellis}},
  \emph{\bibinfo{title}{The large scale structure of spacetime}}
  (\bibinfo{publisher}{Cambridge University Press},
  \bibinfo{address}{Cambridge, England}, \bibinfo{year}{1973}), ISBN
  \bibinfo{isbn}{0-521-09906-4}.

\bibitem[{\citenamefont{Kriele and Hayward}(1997)}]{Kriele97}
\bibinfo{author}{\bibfnamefont{M.}~\bibnamefont{Kriele}} \bibnamefont{and}
  \bibinfo{author}{\bibfnamefont{S.~A.} \bibnamefont{Hayward}},
  \emph{\bibinfo{title}{Outer trapped surfaces and their apparent horizon}},
  \bibinfo{journal}{J. Math. Phys.} \textbf{\bibinfo{volume}{38}},
  \bibinfo{pages}{1593} (\bibinfo{year}{1997}).

\bibitem[{\citenamefont{Andersson et~al.}(2005)\citenamefont{Andersson, Mars,
  and Simon}}]{ams05}
\bibinfo{author}{\bibfnamefont{L.}~\bibnamefont{Andersson}},
  \bibinfo{author}{\bibfnamefont{M.}~\bibnamefont{Mars}}, \bibnamefont{and}
  \bibinfo{author}{\bibfnamefont{W.}~\bibnamefont{Simon}},
  \emph{\bibinfo{title}{Local existence of dynamical and trapping horizons}},
  \bibinfo{journal}{Phys. Rev. Lett.} \textbf{\bibinfo{volume}{95}},
  \bibinfo{pages}{111102} (\bibinfo{year}{2005}), \eprint{gr-qc/0506013}.

\bibitem[{\citenamefont{Dreyer et~al.}(2003)\citenamefont{Dreyer, Krishnan,
  Shoemaker, and Schnetter}}]{Dreyer02a}
\bibinfo{author}{\bibfnamefont{O.}~\bibnamefont{Dreyer}},
  \bibinfo{author}{\bibfnamefont{B.}~\bibnamefont{Krishnan}},
  \bibinfo{author}{\bibfnamefont{D.}~\bibnamefont{Shoemaker}},
  \bibnamefont{and}
  \bibinfo{author}{\bibfnamefont{E.}~\bibnamefont{Schnetter}},
  \emph{\bibinfo{title}{Introduction to {Isolated} {Horizons} in {Numerical}
  {Relativity}}}, \bibinfo{journal}{Phys. Rev. D}
  \textbf{\bibinfo{volume}{67}}, \bibinfo{pages}{024018}
  (\bibinfo{year}{2003}), \eprint{gr-qc/0206008},
  \urlprefix\url{http://link.aps.org/abstract/PRD/v67/e024018}.

\bibitem[{\citenamefont{Dain et~al.}(2005)\citenamefont{Dain, Jaramillo, and
  Krishnan}}]{Dain-2005}
\bibinfo{author}{\bibfnamefont{S.}~\bibnamefont{Dain}},
  \bibinfo{author}{\bibfnamefont{J.~L.} \bibnamefont{Jaramillo}},
  \bibnamefont{and} \bibinfo{author}{\bibfnamefont{B.}~\bibnamefont{Krishnan}},
  \emph{\bibinfo{title}{On the existence of initial data containing isolated
  black holes}}, \bibinfo{journal}{Phys. Rev. D} \textbf{\bibinfo{volume}{71}},
  \bibinfo{pages}{064003} (\bibinfo{year}{2005}), \eprint{gr-qc/0412061}.

\bibitem[{\citenamefont{Jaramillo et~al.}(2004)\citenamefont{Jaramillo,
  Gourgoulhon, and Mena~Marugan}}]{Jaramillo:2004uc}
\bibinfo{author}{\bibfnamefont{J.~L.} \bibnamefont{Jaramillo}},
  \bibinfo{author}{\bibfnamefont{E.}~\bibnamefont{Gourgoulhon}},
  \bibnamefont{and} \bibinfo{author}{\bibfnamefont{G.~A.}
  \bibnamefont{Mena~Marugan}}, \emph{\bibinfo{title}{Inner boundary conditions
  for black hole initial data derived from isolated horizons}},
  \bibinfo{journal}{Phys. Rev. D} \textbf{\bibinfo{volume}{70}},
  \bibinfo{pages}{124036} (\bibinfo{year}{2004}),
  \bibinfo{note}{gr-qc/0407063}.

\bibitem[{\citenamefont{Cook and Pfeiffer}(2004)}]{Cook:2004kt}
\bibinfo{author}{\bibfnamefont{G.~B.} \bibnamefont{Cook}} \bibnamefont{and}
  \bibinfo{author}{\bibfnamefont{H.~P.} \bibnamefont{Pfeiffer}},
  \emph{\bibinfo{title}{Excision boundary conditions for black hole initial
  data}}, \bibinfo{journal}{Phys. Rev. D} \textbf{\bibinfo{volume}{70}}
  (\bibinfo{year}{2004}), \eprint{gr-qc/0407078}.

\bibitem[{\citenamefont{Caudill et~al.}()\citenamefont{Caudill, Cook, Grigsby,
  and Pfeiffer}}]{Cook-2006}
\bibinfo{author}{\bibfnamefont{M.}~\bibnamefont{Caudill}},
  \bibinfo{author}{\bibfnamefont{G.~B.} \bibnamefont{Cook}},
  \bibinfo{author}{\bibfnamefont{J.~D.} \bibnamefont{Grigsby}},
  \bibnamefont{and} \bibinfo{author}{\bibfnamefont{H.~P.}
  \bibnamefont{Pfeiffer}}, \emph{\bibinfo{title}{Circular orbits and spin in
  black-hole initial data}}, \eprint{gr-qc/0605053}.

\bibitem[{\citenamefont{Ashtekar and Galloway}(2005)}]{Ashtekar05}
\bibinfo{author}{\bibfnamefont{A.}~\bibnamefont{Ashtekar}} \bibnamefont{and}
  \bibinfo{author}{\bibfnamefont{G.}~\bibnamefont{Galloway}},
  \emph{\bibinfo{title}{Some uniqueness results for dynamical horizons}},
  \bibinfo{journal}{Advances in Theoretical and Mathematical Physics}
  \textbf{\bibinfo{volume}{to appear}} (\bibinfo{year}{2005}),
  \eprint{gr-qc/0503109}.

\bibitem[{\citenamefont{Booth and Fairhurst}(2004)}]{Booth04a}
\bibinfo{author}{\bibfnamefont{I.}~\bibnamefont{Booth}} \bibnamefont{and}
  \bibinfo{author}{\bibfnamefont{S.}~\bibnamefont{Fairhurst}},
  \emph{\bibinfo{title}{The first law for slowly evolving horizons}},
  \bibinfo{journal}{Phys. Rev. Lett} \textbf{\bibinfo{volume}{92}},
  \bibinfo{pages}{011102} (\bibinfo{year}{2004}), \eprint{gr-qc/0307087}.

\bibitem[{\citenamefont{Kavanagh and Booth}(2006)}]{Kavanagh06}
\bibinfo{author}{\bibfnamefont{W.}~\bibnamefont{Kavanagh}} \bibnamefont{and}
  \bibinfo{author}{\bibfnamefont{I.}~\bibnamefont{Booth}},
  \emph{\bibinfo{title}{Spacetimes containing slowly evolving horizons}}
  (\bibinfo{year}{2006}), \bibinfo{note}{gr-qc/0603074}.

\bibitem[{\citenamefont{Booth and Fairhurst}(2005)}]{Booth05}
\bibinfo{author}{\bibfnamefont{I.}~\bibnamefont{Booth}} \bibnamefont{and}
  \bibinfo{author}{\bibfnamefont{S.}~\bibnamefont{Fairhurst}},
  \emph{\bibinfo{title}{Horizon energy and angular momentum from a hamiltonian
  perspective}}, \bibinfo{journal}{Class. Quant. Grav.}
  \textbf{\bibinfo{volume}{22}}, \bibinfo{pages}{4515} (\bibinfo{year}{2005}),
  \eprint{gr-qc/0505049}.

\bibitem[{\citenamefont{Vaidya}(1951)}]{Vaidya51a}
\bibinfo{author}{\bibfnamefont{P.~C.} \bibnamefont{Vaidya}},
  \emph{\bibinfo{title}{The gravitational field of a radiating star}},
  \bibinfo{journal}{Proc. Ind. Acad. Sci. A} \textbf{\bibinfo{volume}{33}},
  \bibinfo{pages}{264} (\bibinfo{year}{1951}).

\bibitem[{\citenamefont{Kuroda}(1984)}]{Kuroda84a}
\bibinfo{author}{\bibfnamefont{Y.}~\bibnamefont{Kuroda}},
  \emph{\bibinfo{title}{Naked singularities in the {V}aidya spacetime}},
  \bibinfo{journal}{Prog. Theor. Phys.} \textbf{\bibinfo{volume}{72}},
  \bibinfo{pages}{63} (\bibinfo{year}{1984}).

\bibitem[{\citenamefont{Schnetter and
  Krishnan}(2006)}]{Schnetter-Krishnan-2005}
\bibinfo{author}{\bibfnamefont{E.}~\bibnamefont{Schnetter}} \bibnamefont{and}
  \bibinfo{author}{\bibfnamefont{B.}~\bibnamefont{Krishnan}},
  \emph{\bibinfo{title}{Non-symmetric trapped surfaces in the {S}chwarzschild
  and {V}aidya spacetimes}}, \bibinfo{journal}{Phys. Rev. D}
  \textbf{\bibinfo{volume}{73}}, \bibinfo{pages}{021502(R)}
  (\bibinfo{year}{2006}), \eprint{gr-qc/0511017}.

\bibitem[{\citenamefont{Booth et~al.}(2006)\citenamefont{Booth, Brits,
  Gonzalez, and Broeck}}]{Booth05a}
\bibinfo{author}{\bibfnamefont{I.}~\bibnamefont{Booth}},
  \bibinfo{author}{\bibfnamefont{L.}~\bibnamefont{Brits}},
  \bibinfo{author}{\bibfnamefont{J.~A.} \bibnamefont{Gonzalez}},
  \bibnamefont{and} \bibinfo{author}{\bibfnamefont{C.~V.~D.}
  \bibnamefont{Broeck}}, \emph{\bibinfo{title}{Marginally trapped tubes and
  dynamical horizons}}, \bibinfo{journal}{Class. Quant. Grav.}
  \textbf{\bibinfo{volume}{23}}, \bibinfo{pages}{413} (\bibinfo{year}{2006}),
  \eprint{gr-qc/0506119}.

\bibitem[{\citenamefont{Thornburg}(2004)}]{Thornburg2003:AH-finding}
\bibinfo{author}{\bibfnamefont{J.}~\bibnamefont{Thornburg}},
  \emph{\bibinfo{title}{A fast apparent-horizon finder for 3-dimensional
  {C}artesian grids in numerical relativity}}, \bibinfo{journal}{Class. Quantum
  Grav.} \textbf{\bibinfo{volume}{21}}, \bibinfo{pages}{743}
  (\bibinfo{year}{2004}), \eprint{gr-qc/0306056},
  \urlprefix\url{http://stacks.iop.org/0264-9381/21/743}.

\bibitem[{\citenamefont{Eardley}(1998)}]{Eardley98}
\bibinfo{author}{\bibfnamefont{D.~M.} \bibnamefont{Eardley}},
  \emph{\bibinfo{title}{Black hole boundary conditions and coordinate
  conditions}}, \bibinfo{journal}{Phys. Rev. D} \textbf{\bibinfo{volume}{57}},
  \bibinfo{pages}{2299} (\bibinfo{year}{1998}).

\bibitem[{\citenamefont{Ben-Dov}(2004)}]{BenDov04a}
\bibinfo{author}{\bibfnamefont{I.}~\bibnamefont{Ben-Dov}},
  \emph{\bibinfo{title}{The penrose inequality and apparent horizons}},
  \bibinfo{journal}{Phys. Rev. D} \textbf{\bibinfo{volume}{70}},
  \bibinfo{pages}{124031} (\bibinfo{year}{2004}), \eprint{gr-qc/0408066}.

\bibitem[{\citenamefont{Oppenheimer and Snyder}(1939)}]{Oppenheimer39a}
\bibinfo{author}{\bibfnamefont{J.~R.} \bibnamefont{Oppenheimer}}
  \bibnamefont{and} \bibinfo{author}{\bibfnamefont{H.}~\bibnamefont{Snyder}},
  \emph{\bibinfo{title}{On continued gravitational contraction}},
  \bibinfo{journal}{Phys. Rev. D} \textbf{\bibinfo{volume}{56}},
  \bibinfo{pages}{455} (\bibinfo{year}{1939}).

\bibitem[{\citenamefont{Schoen}(2005)}]{Schoen04}
\bibinfo{author}{\bibfnamefont{R.}~\bibnamefont{Schoen}}
  (\bibinfo{year}{2005}), \bibinfo{note}{presentation at the Miami waves
  conference}.

\bibitem[{\citenamefont{Ashtekar et~al.}(2004)\citenamefont{Ashtekar, Engle,
  Pawlowski, and {Van Den Broeck}}}]{Ashtekar04a}
\bibinfo{author}{\bibfnamefont{A.}~\bibnamefont{Ashtekar}},
  \bibinfo{author}{\bibfnamefont{J.}~\bibnamefont{Engle}},
  \bibinfo{author}{\bibfnamefont{T.}~\bibnamefont{Pawlowski}},
  \bibnamefont{and} \bibinfo{author}{\bibfnamefont{C.}~\bibnamefont{{Van Den
  Broeck}}}, \emph{\bibinfo{title}{Multipole moments of isolated horizons}},
  \bibinfo{journal}{Class. Quantum Grav.} \textbf{\bibinfo{volume}{21}},
  \bibinfo{pages}{2549} (\bibinfo{year}{2004}), \eprint{gr-qc/0401114}.

\bibitem[{\citenamefont{Brill and Lindquist}(1963)}]{Brill63}
\bibinfo{author}{\bibfnamefont{D.~S.} \bibnamefont{Brill}} \bibnamefont{and}
  \bibinfo{author}{\bibfnamefont{R.~W.} \bibnamefont{Lindquist}},
  \emph{\bibinfo{title}{Interaction energy in geometrostatics}},
  \bibinfo{journal}{Phys. Rev.} \textbf{\bibinfo{volume}{131}},
  \bibinfo{pages}{471} (\bibinfo{year}{1963}).

\bibitem[{\citenamefont{Cook}(2000)}]{Cook00a1}
\bibinfo{author}{\bibfnamefont{G.~B.} \bibnamefont{Cook}},
  \emph{\bibinfo{title}{Initial data for numerical relativity}},
  \bibinfo{journal}{Living Rev. Rel.} \textbf{\bibinfo{volume}{3}},
  \bibinfo{pages}{5} (\bibinfo{year}{2000}),
  \urlprefix\url{http://relativity.livingreviews.org/Articles/lrr-2000-5/index%
.html}.

\bibitem[{\citenamefont{Misner}(1963)}]{Misner63}
\bibinfo{author}{\bibfnamefont{C.~W.} \bibnamefont{Misner}},
  \emph{\bibinfo{title}{The method of images in geometrostatics}},
  \bibinfo{journal}{Ann. Phys.} \textbf{\bibinfo{volume}{24}},
  \bibinfo{pages}{102} (\bibinfo{year}{1963}).

\bibitem[{\citenamefont{Krishnan}(2002)}]{Krishnan02}
\bibinfo{author}{\bibfnamefont{B.}~\bibnamefont{Krishnan}},
  \emph{\bibinfo{title}{Isolated horizons in numerical relativity}}, Ph.D.
  thesis, \bibinfo{school}{Pennsylvania State University}
  (\bibinfo{year}{2002}),
  \urlprefix\url{http://etda.libraries.psu.edu/theses/approved/WorldWideIndex/%
ETD-177/index.html}.

\bibitem[{\citenamefont{Alcubierre and Br\"ugmann}(2001)}]{Alcubierre00a}
\bibinfo{author}{\bibfnamefont{M.}~\bibnamefont{Alcubierre}} \bibnamefont{and}
  \bibinfo{author}{\bibfnamefont{B.}~\bibnamefont{Br\"ugmann}},
  \emph{\bibinfo{title}{Simple excision of a black hole in 3+1 numerical
  relativity}}, \bibinfo{journal}{Phys. Rev. D} \textbf{\bibinfo{volume}{63}},
  \bibinfo{pages}{104006} (\bibinfo{year}{2001}), \eprint{gr-qc/0008067}.

\bibitem[{\citenamefont{Alcubierre et~al.}(2001)\citenamefont{Alcubierre,
  Br\"ugmann, Pollney, Seidel, and Takahashi}}]{Alcubierre01a}
\bibinfo{author}{\bibfnamefont{M.}~\bibnamefont{Alcubierre}},
  \bibinfo{author}{\bibfnamefont{B.}~\bibnamefont{Br\"ugmann}},
  \bibinfo{author}{\bibfnamefont{D.}~\bibnamefont{Pollney}},
  \bibinfo{author}{\bibfnamefont{E.}~\bibnamefont{Seidel}}, \bibnamefont{and}
  \bibinfo{author}{\bibfnamefont{R.}~\bibnamefont{Takahashi}},
  \emph{\bibinfo{title}{Black hole excision for dynamic black holes}},
  \bibinfo{journal}{Phys. Rev. D} \textbf{\bibinfo{volume}{64}},
  \bibinfo{pages}{061501(R)} (\bibinfo{year}{2001}), \eprint{gr-qc/0104020}.

\bibitem[{\citenamefont{Alcubierre et~al.}(2003)\citenamefont{Alcubierre,
  Br\"ugmann, Diener, Koppitz, Pollney, Seidel, and Takahashi}}]{Alcubierre02a}
\bibinfo{author}{\bibfnamefont{M.}~\bibnamefont{Alcubierre}},
  \bibinfo{author}{\bibfnamefont{B.}~\bibnamefont{Br\"ugmann}},
  \bibinfo{author}{\bibfnamefont{P.}~\bibnamefont{Diener}},
  \bibinfo{author}{\bibfnamefont{M.}~\bibnamefont{Koppitz}},
  \bibinfo{author}{\bibfnamefont{D.}~\bibnamefont{Pollney}},
  \bibinfo{author}{\bibfnamefont{E.}~\bibnamefont{Seidel}}, \bibnamefont{and}
  \bibinfo{author}{\bibfnamefont{R.}~\bibnamefont{Takahashi}},
  \emph{\bibinfo{title}{Gauge conditions for long-term numerical black hole
  evolutions without excision}}, \bibinfo{journal}{Phys. Rev. D}
  \textbf{\bibinfo{volume}{67}}, \bibinfo{pages}{084023}
  (\bibinfo{year}{2003}), \eprint{gr-qc/0206072}.

\bibitem[{\citenamefont{Goodale et~al.}(2003)\citenamefont{Goodale, Allen,
  Lanfermann, Mass\'o, Radke, Seidel, and Shalf}}]{Goodale02a}
\bibinfo{author}{\bibfnamefont{T.}~\bibnamefont{Goodale}},
  \bibinfo{author}{\bibfnamefont{G.}~\bibnamefont{Allen}},
  \bibinfo{author}{\bibfnamefont{G.}~\bibnamefont{Lanfermann}},
  \bibinfo{author}{\bibfnamefont{J.}~\bibnamefont{Mass\'o}},
  \bibinfo{author}{\bibfnamefont{T.}~\bibnamefont{Radke}},
  \bibinfo{author}{\bibfnamefont{E.}~\bibnamefont{Seidel}}, \bibnamefont{and}
  \bibinfo{author}{\bibfnamefont{J.}~\bibnamefont{Shalf}},
  \emph{\bibinfo{title}{The {C}actus framework and toolkit: Design and
  applications}}, in \emph{\bibinfo{booktitle}{Vector and Parallel Processing
  -- VECPAR'2002, 5th International Conference, Lecture Notes in Computer
  Science}} (\bibinfo{publisher}{Springer}, \bibinfo{address}{Berlin},
  \bibinfo{year}{2003}),
  \urlprefix\url{http://www.cactuscode.org/Publications/}.

\bibitem[{cac({\natexlab{a}})}]{cactusweb1}
\bibinfo{note}{{Cactus} Computational Toolkit home page},
  \urlprefix\url{http://www.cactuscode.org/}.

\bibitem[{\citenamefont{Schnetter et~al.}(2004)\citenamefont{Schnetter, Hawley,
  and Hawke}}]{Schnetter-etal-03b}
\bibinfo{author}{\bibfnamefont{E.}~\bibnamefont{Schnetter}},
  \bibinfo{author}{\bibfnamefont{S.~H.} \bibnamefont{Hawley}},
  \bibnamefont{and} \bibinfo{author}{\bibfnamefont{I.}~\bibnamefont{Hawke}},
  \emph{\bibinfo{title}{Evolutions in {3D} numerical relativity using fixed
  mesh refinement}}, \bibinfo{journal}{Class. Quantum Grav.}
  \textbf{\bibinfo{volume}{21}}, \bibinfo{pages}{1465} (\bibinfo{year}{2004}),
  \eprint{gr-qc/0310042}.

\bibitem[{car()}]{carpetweb}
\bibinfo{note}{Mesh Refinement with Carpet},
  \urlprefix\url{http://www.carpetcode.org/}.

\bibitem[{\citenamefont{Thornburg}(1996)}]{Thornburg95}
\bibinfo{author}{\bibfnamefont{J.}~\bibnamefont{Thornburg}},
  \emph{\bibinfo{title}{Finding apparent horizons in numerical relativity}},
  \bibinfo{journal}{Phys. Rev. D} \textbf{\bibinfo{volume}{54}},
  \bibinfo{pages}{4899} (\bibinfo{year}{1996}), \eprint{gr-qc/9508014}.

\bibitem[{\citenamefont{Brandt and Br{\"u}gmann}(1997)}]{Brandt97b}
\bibinfo{author}{\bibfnamefont{S.}~\bibnamefont{Brandt}} \bibnamefont{and}
  \bibinfo{author}{\bibfnamefont{B.}~\bibnamefont{Br{\"u}gmann}},
  \emph{\bibinfo{title}{A simple construction of initial data for multiple
  black holes}}, \bibinfo{journal}{Phys. Rev. Lett.}
  \textbf{\bibinfo{volume}{78}}, \bibinfo{pages}{3606} (\bibinfo{year}{1997}),
  \eprint{gr-qc/9703066}.

\bibitem[{\citenamefont{Cook}(1994)}]{Cook94}
\bibinfo{author}{\bibfnamefont{G.~B.} \bibnamefont{Cook}},
  \emph{\bibinfo{title}{Three-dimensional initial data for the collision of two
  black holes {II}: {Q}uasi-circular orbits for equal-mass black holes}},
  \bibinfo{journal}{Phys. Rev. D} \textbf{\bibinfo{volume}{50}},
  \bibinfo{pages}{5025} (\bibinfo{year}{1994}).

\bibitem[{\citenamefont{Baker et~al.}(2002)\citenamefont{Baker, Campanelli,
  Lousto, and Takahashi}}]{Baker:2002qf}
\bibinfo{author}{\bibfnamefont{J.}~\bibnamefont{Baker}},
  \bibinfo{author}{\bibfnamefont{M.}~\bibnamefont{Campanelli}},
  \bibinfo{author}{\bibfnamefont{C.~O.} \bibnamefont{Lousto}},
  \bibnamefont{and}
  \bibinfo{author}{\bibfnamefont{R.}~\bibnamefont{Takahashi}},
  \emph{\bibinfo{title}{Modeling gravitational radiation from coalescing binary
  black holes}}, \bibinfo{journal}{Phys. Rev. D} \textbf{\bibinfo{volume}{65}},
  \bibinfo{pages}{124012} (\bibinfo{year}{2002}),
  \eprint[http://arXiv.org/abs]{astro-ph/0202469}.

\bibitem[{\citenamefont{Baker et~al.}(2003)\citenamefont{Baker, Campanelli,
  Lousto, and Takahashi}}]{Baker:2003ds}
\bibinfo{author}{\bibfnamefont{J.}~\bibnamefont{Baker}},
  \bibinfo{author}{\bibfnamefont{M.}~\bibnamefont{Campanelli}},
  \bibinfo{author}{\bibfnamefont{C.~O.} \bibnamefont{Lousto}},
  \bibnamefont{and}
  \bibinfo{author}{\bibfnamefont{R.}~\bibnamefont{Takahashi}},
  \emph{\bibinfo{title}{The final plunge of spinning binary black holes}}
  (\bibinfo{year}{2003}), \eprint[http://arXiv.org/abs]{astro-ph/0305287}.

\bibitem[{\citenamefont{Alcubierre et~al.}(2005)\citenamefont{Alcubierre,
  Br{\"u}gmann, Diener, Guzm{\'a}n, Hawke, Hawley, Herrmann, Koppitz, Pollney,
  Seidel et~al.}}]{Alcubierre2003:pre-ISCO-coalescence-times}
\bibinfo{author}{\bibfnamefont{M.}~\bibnamefont{Alcubierre}},
  \bibinfo{author}{\bibfnamefont{B.}~\bibnamefont{Br{\"u}gmann}},
  \bibinfo{author}{\bibfnamefont{P.}~\bibnamefont{Diener}},
  \bibinfo{author}{\bibfnamefont{F.~S.} \bibnamefont{Guzm{\'a}n}},
  \bibinfo{author}{\bibfnamefont{I.}~\bibnamefont{Hawke}},
  \bibinfo{author}{\bibfnamefont{S.}~\bibnamefont{Hawley}},
  \bibinfo{author}{\bibfnamefont{F.}~\bibnamefont{Herrmann}},
  \bibinfo{author}{\bibfnamefont{M.}~\bibnamefont{Koppitz}},
  \bibinfo{author}{\bibfnamefont{D.}~\bibnamefont{Pollney}},
  \bibinfo{author}{\bibfnamefont{E.}~\bibnamefont{Seidel}},
  \bibnamefont{et~al.}, \emph{\bibinfo{title}{Dynamical evolution of
  quasi-circular binary black hole data}}, \bibinfo{journal}{Phys. Rev. D}
  \textbf{\bibinfo{volume}{72}}, \bibinfo{pages}{044004}
  (\bibinfo{year}{2005}), \eprint{gr-qc/0411149},
  \urlprefix\url{http://link.aps.org/abstract/PRD/v72/e044004}.

\bibitem[{\citenamefont{Br\"ugmann et~al.}(2004)\citenamefont{Br\"ugmann,
  Tichy, and Jansen}}]{Bruegmann:2003aw}
\bibinfo{author}{\bibfnamefont{B.}~\bibnamefont{Br\"ugmann}},
  \bibinfo{author}{\bibfnamefont{W.}~\bibnamefont{Tichy}}, \bibnamefont{and}
  \bibinfo{author}{\bibfnamefont{N.}~\bibnamefont{Jansen}},
  \emph{\bibinfo{title}{Numerical simulation of orbiting black holes}},
  \bibinfo{journal}{Phys. Rev. Lett.} \textbf{\bibinfo{volume}{92}},
  \bibinfo{pages}{211101} (\bibinfo{year}{2004}), \eprint{gr-qc/0312112}.

\bibitem[{\citenamefont{Alcubierre et~al.}(2006)\citenamefont{Alcubierre,
  Diener, Guzm\'an, Hawley, Koppitz, Pollney, and
  Seidel}}]{Alcubierre2003:co-rotating-shift}
\bibinfo{author}{\bibfnamefont{M.}~\bibnamefont{Alcubierre}},
  \bibinfo{author}{\bibfnamefont{P.}~\bibnamefont{Diener}},
  \bibinfo{author}{\bibfnamefont{F.~S.} \bibnamefont{Guzm\'an}},
  \bibinfo{author}{\bibfnamefont{S.}~\bibnamefont{Hawley}},
  \bibinfo{author}{\bibfnamefont{M.}~\bibnamefont{Koppitz}},
  \bibinfo{author}{\bibfnamefont{D.}~\bibnamefont{Pollney}}, \bibnamefont{and}
  \bibinfo{author}{\bibfnamefont{E.}~\bibnamefont{Seidel}},
  \emph{\bibinfo{title}{Shift {C}onditions for {O}rbiting {B}inaries in
  {N}umerical {R}elativity}} (\bibinfo{year}{2006}), \bibinfo{note}{in
  preparation}.

\bibitem[{\citenamefont{Ansorg et~al.}(2004)\citenamefont{Ansorg, Br\"ugmann,
  and Tichy}}]{Ansorg:2004ds}
\bibinfo{author}{\bibfnamefont{M.}~\bibnamefont{Ansorg}},
  \bibinfo{author}{\bibfnamefont{B.}~\bibnamefont{Br\"ugmann}},
  \bibnamefont{and} \bibinfo{author}{\bibfnamefont{W.}~\bibnamefont{Tichy}},
  \emph{\bibinfo{title}{A single-domain spectral method for black hole puncture
  data}}, \bibinfo{journal}{Phys. Rev. D} \textbf{\bibinfo{volume}{70}},
  \bibinfo{pages}{064011} (\bibinfo{year}{2004}), \eprint{gr-qc/0404056}.

\bibitem[{\citenamefont{Schnetter et~al.}(2005)\citenamefont{Schnetter,
  Herrmann, and Pollney}}]{Schnetter04}
\bibinfo{author}{\bibfnamefont{E.}~\bibnamefont{Schnetter}},
  \bibinfo{author}{\bibfnamefont{F.}~\bibnamefont{Herrmann}}, \bibnamefont{and}
  \bibinfo{author}{\bibfnamefont{D.}~\bibnamefont{Pollney}},
  \emph{\bibinfo{title}{Horizon pretracking}}, \bibinfo{journal}{Phys. Rev. D}
  \textbf{\bibinfo{volume}{71}}, \bibinfo{pages}{044033}
  (\bibinfo{year}{2005}), \eprint{gr-qc/0410081}.

\bibitem[{\citenamefont{Baiotti
  et~al.}(2005{\natexlab{a}})\citenamefont{Baiotti, Hawke, Montero,
  L{\"o}ffler, Rezzolla, Stergioulas, Font, and Seidel}}]{Baiotti04}
\bibinfo{author}{\bibfnamefont{L.}~\bibnamefont{Baiotti}},
  \bibinfo{author}{\bibfnamefont{I.}~\bibnamefont{Hawke}},
  \bibinfo{author}{\bibfnamefont{P.~J.} \bibnamefont{Montero}},
  \bibinfo{author}{\bibfnamefont{F.}~\bibnamefont{L{\"o}ffler}},
  \bibinfo{author}{\bibfnamefont{L.}~\bibnamefont{Rezzolla}},
  \bibinfo{author}{\bibfnamefont{N.}~\bibnamefont{Stergioulas}},
  \bibinfo{author}{\bibfnamefont{J.~A.} \bibnamefont{Font}}, \bibnamefont{and}
  \bibinfo{author}{\bibfnamefont{E.}~\bibnamefont{Seidel}},
  \emph{\bibinfo{title}{Three-dimensional relativistic simulations of rotating
  neutron star collapse to a kerr black hole}}, \bibinfo{journal}{Phys. Rev. D}
  \textbf{\bibinfo{volume}{71}}, \bibinfo{pages}{024035}
  (\bibinfo{year}{2005}{\natexlab{a}}), \eprint{gr-qc/0403029}.

\bibitem[{\citenamefont{Baiotti
  et~al.}(2005{\natexlab{b}})\citenamefont{Baiotti, Hawke, Rezzolla, and
  Schnetter}}]{Baiotti04b}
\bibinfo{author}{\bibfnamefont{L.}~\bibnamefont{Baiotti}},
  \bibinfo{author}{\bibfnamefont{I.}~\bibnamefont{Hawke}},
  \bibinfo{author}{\bibfnamefont{L.}~\bibnamefont{Rezzolla}}, \bibnamefont{and}
  \bibinfo{author}{\bibfnamefont{E.}~\bibnamefont{Schnetter}},
  \emph{\bibinfo{title}{Gravitational-wave emission from rotating gravitational
  collapse in three dimensions}}, \bibinfo{journal}{Phys. Rev. Lett.}
  \textbf{\bibinfo{volume}{94}}, \bibinfo{pages}{131101}
  (\bibinfo{year}{2005}{\natexlab{b}}), \eprint{gr-qc/0503016}.

\bibitem[{\citenamefont{Ott et~al.}(2006)\citenamefont{Ott, Dimmelmeier, Hawke,
  Schnetter, Zink, M{\"u}ller, and Seidel}}]{Ott06a}
\bibinfo{author}{\bibfnamefont{C.~D.} \bibnamefont{Ott}},
  \bibinfo{author}{\bibfnamefont{H.}~\bibnamefont{Dimmelmeier}},
  \bibinfo{author}{\bibfnamefont{I.}~\bibnamefont{Hawke}},
  \bibinfo{author}{\bibfnamefont{E.}~\bibnamefont{Schnetter}},
  \bibinfo{author}{\bibfnamefont{B.}~\bibnamefont{Zink}},
  \bibinfo{author}{\bibfnamefont{E.}~\bibnamefont{M{\"u}ller}},
  \bibnamefont{and} \bibinfo{author}{\bibfnamefont{E.}~\bibnamefont{Seidel}},
  \emph{\bibinfo{title}{Fully consistent {3D} general relativistic rotating
  stellar core collapse with mesh refinement: Comparison to {2D} {CFC}-based
  approach}} (\bibinfo{year}{2006}), \bibinfo{note}{in preparation}.

\bibitem[{\citenamefont{Kreiss and Oliger}(1973)}]{Kreiss73}
\bibinfo{author}{\bibfnamefont{H.-O.} \bibnamefont{Kreiss}} \bibnamefont{and}
  \bibinfo{author}{\bibfnamefont{J.}~\bibnamefont{Oliger}},
  \emph{\bibinfo{title}{Methods for the approximate solution of time dependent
  problems}}, \bibinfo{journal}{Global atmospheric research programme
  publications series} \textbf{\bibinfo{volume}{10}} (\bibinfo{year}{1973}).

\bibitem[{\citenamefont{Price}(1972)}]{Price72}
\bibinfo{author}{\bibfnamefont{R.}~\bibnamefont{Price}},
  \emph{\bibinfo{title}{Nonspherical perturbations of relativistic
  gravitational collapse. {I}. scalar and gravitational perturtbations}},
  \bibinfo{journal}{Phys. Rev. D} \textbf{\bibinfo{volume}{5}},
  \bibinfo{pages}{2419} (\bibinfo{year}{1972}).

\bibitem[{\citenamefont{Dafermos and Rodnianski}(2005)}]{Dafermos05}
\bibinfo{author}{\bibfnamefont{M.}~\bibnamefont{Dafermos}} \bibnamefont{and}
  \bibinfo{author}{\bibfnamefont{I.}~\bibnamefont{Rodnianski}},
  \emph{\bibinfo{title}{A proof of price's law for the collapse of a
  self-gravitating scalar field}}, \bibinfo{journal}{Invent. Math.}
  \textbf{\bibinfo{volume}{162}}, \bibinfo{pages}{381} (\bibinfo{year}{2005}),
  \eprint{gr-qc/0309115}.

\bibitem[{\citenamefont{Anninos
  et~al.}(1995{\natexlab{b}})\citenamefont{Anninos, Daues, Mass{\'o}, Seidel,
  and Suen}}]{Anninos94e}
\bibinfo{author}{\bibfnamefont{P.}~\bibnamefont{Anninos}},
  \bibinfo{author}{\bibfnamefont{G.}~\bibnamefont{Daues}},
  \bibinfo{author}{\bibfnamefont{J.}~\bibnamefont{Mass{\'o}}},
  \bibinfo{author}{\bibfnamefont{E.}~\bibnamefont{Seidel}}, \bibnamefont{and}
  \bibinfo{author}{\bibfnamefont{W.-M.} \bibnamefont{Suen}},
  \emph{\bibinfo{title}{Horizon boundary conditions for black hole
  spacetimes}}, \bibinfo{journal}{Phys. Rev. D} \textbf{\bibinfo{volume}{51}},
  \bibinfo{pages}{5562} (\bibinfo{year}{1995}{\natexlab{b}}).

\bibitem[{cac({\natexlab{b}})}]{cactuseinsteinweb}
\bibinfo{note}{{CactusEinstein} Toolkit home page},
  \urlprefix\url{http://www.cactuscode.org/Community/numericalRelativity/}.

\bibitem[{whi()}]{whiskyweb}
\bibinfo{note}{{Whisky}, {EU} {Network} {GR} Hydrodynamics Code},
  \urlprefix\url{http://www.whiskycode.org/}.

\bibitem[{\citenamefont{Anderson et~al.}(1999)\citenamefont{Anderson, Bai,
  Bischof, Blackford, Demmel, Dongarra, Du~Croz, Greenbaum, Hammarling,
  McKenney et~al.}}]{laug}
\bibinfo{author}{\bibfnamefont{E.}~\bibnamefont{Anderson}},
  \bibinfo{author}{\bibfnamefont{Z.}~\bibnamefont{Bai}},
  \bibinfo{author}{\bibfnamefont{C.}~\bibnamefont{Bischof}},
  \bibinfo{author}{\bibfnamefont{S.}~\bibnamefont{Blackford}},
  \bibinfo{author}{\bibfnamefont{J.}~\bibnamefont{Demmel}},
  \bibinfo{author}{\bibfnamefont{J.}~\bibnamefont{Dongarra}},
  \bibinfo{author}{\bibfnamefont{J.}~\bibnamefont{Du~Croz}},
  \bibinfo{author}{\bibfnamefont{A.}~\bibnamefont{Greenbaum}},
  \bibinfo{author}{\bibfnamefont{S.}~\bibnamefont{Hammarling}},
  \bibinfo{author}{\bibfnamefont{A.}~\bibnamefont{McKenney}},
  \bibnamefont{et~al.}, \emph{\bibinfo{title}{{LAPACK} Users' Guide}}
  (\bibinfo{publisher}{Society for Industrial and Applied Mathematics},
  \bibinfo{address}{Philadelphia, PA}, \bibinfo{year}{1999}),
  \bibinfo{edition}{3rd} ed., ISBN \bibinfo{isbn}{0-89871-447-8 (paperback)}.

\bibitem[{lap()}]{lapackweb}
\bibinfo{note}{{LAPACK}: Linear Algebra Package},
  \urlprefix\url{http://www.netlib.org/lapack/}.

\bibitem[{bla()}]{blasweb}
\bibinfo{note}{{BLAS}: Basic Linear Algebra Subroutines},
  \urlprefix\url{http://www.netlib.org/blas/}.

\bibitem[{net()}]{netlibweb}
\bibinfo{note}{Netlib Repository}, \urlprefix\url{http://www.netlib.org/}.

\bibitem[{\citenamefont{Press et~al.}(1986)\citenamefont{Press, Flannery,
  Teukolsky, and Vetterling}}]{Press86}
\bibinfo{author}{\bibfnamefont{W.~H.} \bibnamefont{Press}},
  \bibinfo{author}{\bibfnamefont{B.~P.} \bibnamefont{Flannery}},
  \bibinfo{author}{\bibfnamefont{S.~A.} \bibnamefont{Teukolsky}},
  \bibnamefont{and} \bibinfo{author}{\bibfnamefont{W.~T.}
  \bibnamefont{Vetterling}}, \emph{\bibinfo{title}{Numerical Recipes}}
  (\bibinfo{publisher}{Cambridge University Press},
  \bibinfo{address}{Cambridge, England}, \bibinfo{year}{1986}),
  \urlprefix\url{http://www.nr.com/}.

\bibitem[{umf()}]{umfpackweb}
\bibinfo{note}{{UMFPACK}},
  \urlprefix\url{http://www.cise.ufl.edu/research/sparse/umfpack/}.

\end{thebibliography}

\end{document}